\documentclass[]{aa}
\usepackage{aabib99}
\usepackage{graphics} 
\usepackage{epsf} 
\begin{document}
\thesaurus{02(12.12.1; 
	      11.05.2) 
	  }
\title{ Evolution of faint galaxy clustering:}
\subtitle{The 2-point angular correlation function of 20,000 galaxies to 
$V<23.5$ and $I<22.5$.\thanks{based on observations from the Canada-France-
Hawaii Telescope}
}
\author{R\'emi A. Cabanac \inst{1} \thanks{Fellow of Fonds FCAR, Qu\'ebec}
\and Val\'erie de Lapparent \inst{1}
\and Paul Hickson \inst{2}
}
\offprints{cabanac@iap.fr}
\institute{
Institut d'Astrophysique de Paris, CNRS, 98 bis bvd Arago, 75014 Paris 
\and Dept. of Physics \& Astronomy, UBC, 2219 Main Mall, Vancouver, BC V6T1Z4 , Canada 
}
\date{Received ... /Accepted ...}
\titlerunning{$\omega(\theta)$ evolution for 20,000 galaxies...}
\authorrunning{}
\maketitle
\begin{abstract}
The UH8K wide field camera of the CFHT was used to image 0.68 deg$^2$ of sky.
From these images, $\sim$20,000 galaxies were detected to completeness magnitudes 
$V<23.5$ and $I<22.5$. The angular correlation function of these galaxies is
well represented by the parameterization $\omega(\theta) = A_W\,\theta^{-\delta}$. 
The slope $\delta\simeq-0.8$ shows no significant variation over the range of magnitude. 
The amplitude $A_W$ decreases with increasing magnitude in a way that is 
most compatible with a $\Lambda$CDM model ($\Omega_0 = 0.2, \Lambda=0.8$) with 
a hierarchical clustering evolution parameter $\epsilon>0$. We infer a best-fit 
spatial correlation length of $r_{00}\simeq 5.85\pm0.5\,h^{-1}$ Mpc at $z=0$.
The peak redshift of the survey ($I\le22.5$) is estimated to be
$z_{peak}\sim0.58$, using the blue-evolving luminosity function from the CFRS
and the flat $\Lambda$ cosmology, and $r_0(z_{peak})\simeq3.5\pm0.5\,h^{-1}$
Mpc. We also detect a significant difference in clustering amplitude for the
red and blue galaxies, quantitatively measured by correlation
lengths of $r_{00}=5.3\pm0.5\,h^{-1}$ Mpc and $r_{00}=1.9\pm0.9\,h^{-1}$ 
Mpc respectively, at $z=0$.

\end{abstract}
\section{Introduction} \label{intro}
For the past decade, the study of the spatial Large-Scale Structure
(LSS) of the universe has become an major tool for constraining the 
cosmological models. In particular, provided many assumptions on
how morphological type correlates with colour, how mass is correlated
with optical luminosity, and how local density correlates with morphology,
recent CDM hierarchical N-body simulations and semi-analytic models of
galaxy formation are able to make tentative predictions on the
clustering evolution of the galaxies as a function of their redshifts,
spectral types and star formation rates (Kauffmann et al., 1999a; 1999b). 
By measuring redshifts for $10^5$ or more galaxies, the next generation
redshift surveys such as the VIRMOS survey \cite{lefevre98}, the DEEP survey
\cite{davis98}, and the LZT survey (Hickson et al{.}, 1998; 
see also http://www.astro.ubc.ca/LMT/lzt/index.html) will allow detailed studies
of the large-scale clustering and its evolution to $z\sim1$.

Until these surveys are completed, the measurement of the
2-point angular correlation functions $\omega(\theta)$ of large photometric 
galaxy samples remains the best alternative to constrain  galaxy clustering
at $z>1$.  The major caveat of $\omega(\theta)$, as opposed to the 2-point 
spatial correlation function $\xi(r)$, is that it probes the projection of a 3D 
distribution of the galaxies onto the 2D celestial sphere. i.e. one can not 
tell whether a given galaxy is a faint nearby object or a bright remote one. 
As a consequence, $\omega(\theta)$ is sensitive to the effects of both the 
intrinsic 3D clustering and the luminosity evolution (LE) of galaxies for a 
given set of cosmological parameters. 
To avoid this degeneracy, one must choose 
between two approaches to extract sensible information from $\omega(\theta)$. 
First, one may use past observations to assume a scenario of galaxy evolution 
with given LEs and redshift distributions for each galaxy population,
and then deduce the clustering evolution. A second approach would be to assume
a clustering scenario, cosmological parameters, and to measure the
evolution of the correlation function in order to validate the theoretical LE
used to model the galaxy counts, e.g. Roche et al. \cite*{roche93}. 
In this paper, we favour the first approach. We use the Canada-France Redshift 
Survey (CFRS; Lilly et al. 1996) luminosity
function and redshift distribution to invert the angular correlation
function with Limber's formula (cf section \ref{anal} on modeling of
$\omega[\theta]$, $\xi[r]$ and $r_0$) to compute the spatial
correlation length from the amplitude of $\omega(\theta)$. This approach has
several limits which are discussed in section \ref{disc}.

An extensive literature covers the evolution of clustering using 
$\omega(\theta)$. The first attempts to describe clustering using 
counts in cells is due to Limber \cite*{limber54}. The two-point correlation 
function as a statistical diagnosis of clustering has been popularized in 
astrophysics by Hauser \& Peebles \cite*{hauser73} and applied to the Zwicky 
Catalog \cite{peebles74}. Since these pioneering studies, the method, 
fully described by Peebles \cite*{peebles80}, has been applied to 
many photographic catalogues in diverse photometric bands e.g. see Groth \&
Peebles \cite*{groth77}, Koo \& Szalay \cite*{koo84}, Maddox et al.
\cite*{maddox90} where the clustering of local galaxies is studied on large
angular scales. Using Limber's \cite*{limber53} formula relating
$\omega(\theta)$ to the spatial correlation function $\xi(r)$, these studies
establish that the spatial clustering of local galaxies can be parameterized as
a power law, $\xi(r)= (r/r_0)^{-\gamma}$, where the correlation length 
$r_0 \simeq 4-8h^{-1}$ Mpc ($h=H_0/100$) and the slope $\gamma\simeq1.7-1.8$.

A second generation of studies based on small-scale CCDs, probes
smaller areas to deeper magnitudes
\cite{efstathiou91,campos95,neuschaefer92,neuschaefer95b,roukema94,brainerd95,brainerd98,brainerd99,hudon96,lidman96,roche93,roche96,woods97}, therefore allowing to measure the
evolution of the correlation length to redshifts $z\la1$). The most
recent studies take advantage of large area CCD detectors
\cite{roche99,post98} to measure the angular correlation function, and
of the use of photometric redshifts \cite{koo99} to estimate the
spatial correlation function from deep photometric surveys
\cite{villumsen97,arnouts99}.  Despite these numerous studies,
our knowledge of the clustering of galaxies is still rudimentary.  The
main trends are that while a mild luminosity evolution seems to be
required to explain faint number counts in the $I$-band, weak or no
evolution in the galaxy clustering with redshift is detected out to
$z\sim1$.

The existing surveys measuring the galaxy angular correlation
function, and covering a broad range of magnitude bands and limits,
constrain the value of $r_0(z_{peak})$ to the range $1.5-4.5\,h^{-1}$Mpc
for $0.5<z_{peak}<1$.  The dispersion is mainly due to the uncertainty
in our knowledge of the luminosity functions and redshifts
distributions for the different galaxy types at $z\sim1$ (section
\ref{disc}), and possibly to the varying selection biases from survey
to survey.  To be consistent with the values of $r_{00}$ measured from
the nearby redshift surveys and ranging from 4$h^{-1}$Mpc to
8$h^{-1}$Mpc
\cite{lapparent88,loveday92,cole94,tucker97,ratcliffe98,guzzo98}, most
observations of the galaxy clustering favour either constant or
increasing clustering with time in proper coordinates, which is
consistent with N-body simulations of CDM hierarchical universes
\cite{davis85,baugh99,hudon96}.

Note that only a few redshift surveys
allow a direct study of the evolution of the spatial clustering
\cite{lilly95,connolly98,arnouts99,carlberg99}: these survey measure
$r_0(z_{peak})\simeq 1.4-4.5\,h^{-1}$Mpc for $0.5<z_{peak}<1$.  We
underline that except for the results of Carlberg et
al. \cite*{carlberg99}, corresponding to the high value of
$r_0(z_{peak})$, the limited area of the mentioned surveys make them
very sensitive to cosmic variance, and the corresponding results on
the correlation function must be taken with caution.

Moreover, the existing analyses have not yet answered convincingly to
the following two questions: Is there an evolution of the angular
correlation function slope $\delta=\gamma-1$ at faint limiting
magnitudes? And do red-selected objects and blue-selected objects show
a true difference in 3-D clustering?  In addition to providing another
measure of the galaxy clustering at $z\simeq0.5$, the new sample
presented here allows us to address these questions. The paper is
organized as follows, the observations are described in section
\ref{obs}, section \ref{reduc} presents the data reduction, section
\ref{counts} addresses the star/galaxy separations and counts.
Section \ref{anal} details the analysis of the correlation function,
section \ref{res} gives the results, and section \ref{disc} provides a
discussion of our results and a comparison with previous work.

\section{Observations}\label{obs}

\begin{figure*}
\epsfysize=8.0cm
\centerline {\epsfbox[40 250 540 490]{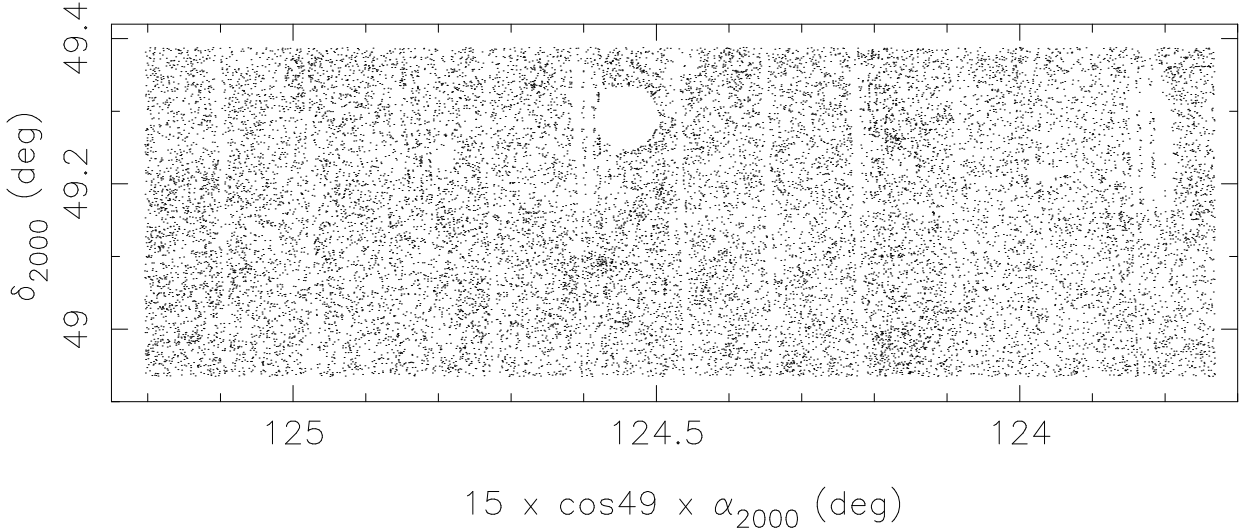}}
\caption{Map of 19,506 galaxies to $I<22.5$.  Bad cosmetics, intervals between
CCD chips, and saturation artifacts due to bright stars are masked (shown
as empty areas).}
\label{fig1}
\end{figure*}

The data were obtained during the spring of 1998 using the prime-focus 
wide-field UH8K CCD Camera \cite{metzger95} at the Canada-France-Hawaii 3.6 m
Telescope. The camera is a mosaic of 8 2k$\times$4k CCD chips covering a total 
area of $\sim28\arcmin\times28\arcmin$ with a scale of $0.206\arcsec$/pixel. 
We observed four fields with R.A. offset of $23\arcmin, 23\arcmin$, and
$17\arcmin$, hence overlapping by 5$\arcmin$, 5$\arcmin$ and 10$\arcmin$
(Table \ref{table1}). The observations were done in bands $V$ (Johnson) and
$I$ (Cousins). The total area of the survey is 0.68 deg$^2$. The limiting
magnitudes are $V<24$ and $I<23$ (see section \ref{completeness} on the
completeness). The central galactic coordinates are $l=126.2^\circ$,
$b=68.2^\circ$. At such high galactic latitudes, the reddening is $E(B-V)<0.03$
mag \cite{burstein82}. This is an upper limit because we observed
in $V$ and $I$ (reddening$\sim\lambda^{-1}$). We neglected both the absolute 
reddening and the relative reddening between the fields. Table \ref{table1} 
gives the characteristics of each field. The seeing is between $0.7\arcsec$ and
$0.8\arcsec$ for both filters during the whole night. The exposure time is
1200 sec for all fields.
Figure \ref{fig1} shows the map of the $\sim$19,500 galaxies detected
to $I<22.5$ (limiting magnitude of the correlation analysis, cf section
\ref{completeness}). In spite of the bad cosmetics of CCD\#4
(visible in the upper right corner of Fig. \ref{fig1}),
the $5\arcmin$ overlaps provide homogeneous sampling of the area,
except for the regions containing bright stars (empty circle in the upper
middle) and the gaps within the UH8K Mosaic CCD chips. 
\begin{table}
\caption{Field characteristics}\label{table1}
\begin{center}
\begin{tabular}{lcccccc}
\hline
Field&$\alpha_{2000}$&$\delta_{2000}$&airmass&airmass\\
&hr~min~sec&$^\circ$~~~$\arcmin$~~~$\arcsec$&in $V$&in $I$\\
\hline
UD02&12~37~55.3&+49~08~50.6&1.491&1.161\\
UD03&12~40~15.6&+49~08~52.3&1.371&1.148\\
UD04&12~42~35.8&+49~08~54.1&1.285&1.154\\
UD05&12~44~06.3&+49~08~55.3&1.224&1.176\\
\hline
\end{tabular}
\end{center}
\noindent
\end{table}
\section{Data reduction}\label{reduc}
The data were reduced using FLIPS by J.C. Cuillandre as part of the
calibration service of the 1998 season of UH8K Camera, Cuillandre 
\cite*{cuillandre98a,cuillandre98b}. 
FLIPS suppresses the dark and bias using the CCD overs-cans and flattens the
response of the 8 detectors using a flat-field made by combining $\sim300$
images of 18 nights of observation with the UH8K Camera.
The final images have a corrected sky flux 
showing variations of less than 1\% within each CCD image. The photometric 
standards were pre-reduced following exactly the same steps. The final dataset
consists of 64 2k$\times$4k frames (32 in $V$, 32 in $I$), and a set of
photometric standard stars.

Photometry was performed using the SExtractor Package \cite{bertin96} which
provides Kron-like elliptical aperture and isophotal fluxes, ($X,Y$)
coordinates, position, elongation and stellarity class for all objects above
a given threshold (we choose a threshold of 25 contiguous pixels above
1.5\,$\sigma$ of sky value in $V$ and $I$). The resulting SExtractor files were
then calibrated for astrometry and photometry.
\subsection{astrometry}\label{astrometry}
\begin{figure}
\epsfysize=12cm
\centerline {\epsfbox[0 0 550 730]{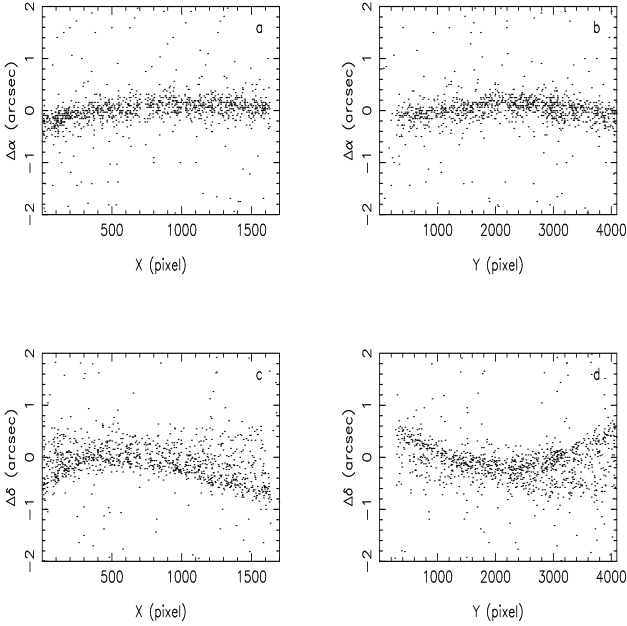}}
\caption{Differences between the equatorial coordinates of the objects
of the overlapping region of the fields UD03-CCD\#0 and UD04-CCD\#6 (in arcsec)
if no correction is applied for the prime focus optical corrector distortion.
(a) shows the right ascension difference $\Delta\alpha$ vs the CCD X axis,
(b) shows $\Delta\alpha$ vs the CCD Y axis, (c) shows the declination difference
$\Delta\delta$ vs the CCD X axis, and (d) shows $\Delta\delta$ vs the CCD Y
axis. Important systematic effects are seen in frame (c) and (d).}
\label{fig2}
\end{figure}

\begin{figure}
\epsfysize=12cm
\centerline {\epsfbox[0 0 550 730]{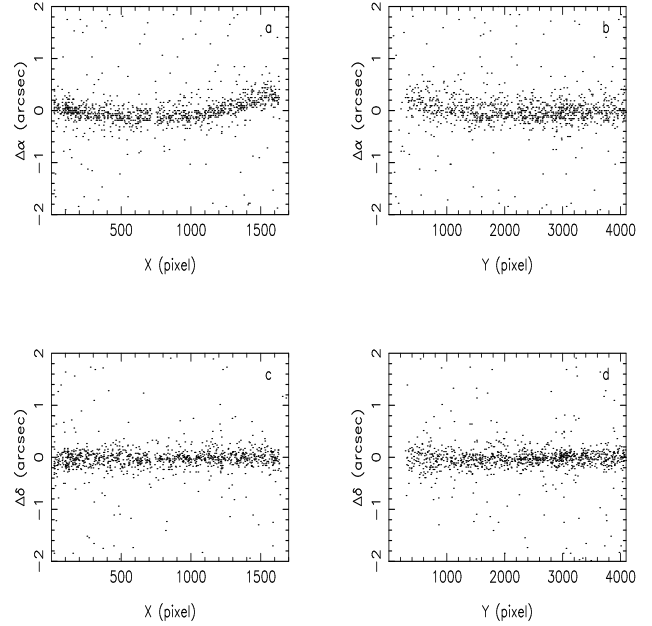}}
\caption{Same as Fig. \protect\ref{fig2} with the correction of the prime focus
optical corrector distortion of eq. (\protect\ref{eq1}).}
\label{fig3}
\end{figure}

The USNO-A2.0 Astrometric Catalogue \cite{monet98} is used as the astrometric
reference as it gives the equatorial coordinates of most objects
in our fields to a red mag$<20$ with accuracies of $\le0.5\arcsec$, and
because it is easily accessible via on-line astronomical databases.
The astrometry is done separately on the 64 frames, with IRAF geomap/geotran
second-order Legendre polynomials. A radial correction is applied previously
on the ($X,Y$) coordinates to correct the prime-focus optical corrector
distortion, whose equations were provided by J.C. Cuillandre
\cite*{cuillandre96}. If $R$ is the actual radial coordinate of an object
from the center of the UH8K mosaic (in mm) and $r$ its observed radial
coordinate (in mm), then the shift $r-R$ due to the corrector is
\begin{eqnarray}\label{eq1} r-R&=&9.07\times10^{-7}~r^3+2.06\times10^{-12}~r^5
\nonumber\\
r&=&0.07284~\theta(1+2.593\times10^{-9}~\theta^{2.093}). \end{eqnarray}
$\theta$ is the angular distance of the object from the center in arcsec. 
According to eq. (\ref{eq1}), the radial distortion is 0.13 pixel at a radius of
3$\arcmin$, 0.32 pixel at 4$\arcmin$, and it becomes non-negligible for radii
greater than 6$\arcmin$ where the distortion is greater then one pixel
(e.g. the radial distortion at 14$\arcmin$ is 14.1 pixels). 
The overlaps between the fields allow us to verify the importance of the
optical distortion correction on the accuracy of the final astrometry.
Figures \ref{fig2} and \ref{fig3} show the errors $\Delta\alpha$ and 
$\Delta\delta$ in X and Y directions for the overlap between the fields UD03
(CCD \#0) and UD04 (CCD \#6) without correction of distortion
(Fig. \ref{fig2}) and with correction (Fig. \ref{fig3}). One can see
that both the systematic errors and the random errors are bellow
0.5$\arcsec$ after correction. We are not able to remove completely the
systematic shifts between the fields (Fig. \ref{fig3} frame a and b).
This may be due to unknown misalignments between the chips of the mosaic.
One can only minimize the systematic effects to sub-arcsecond values.
The final equatorial coordinates are precessed to J2000.

McCracken (priv. comm. 2000) used a finer and more complex approach
with the Canada-France Deep Field (CFDF) ($\sim1$ deg$^2$ covering the CFRS
in UBVRI, the fields in BVRI bands were observed with the UH8K Camera)
by establishing an internal Word-Coordinate System to which all pointing are
referred. The claimed dispersion is 0.3 pix $= 0.06\arcsec$ over the entire
camera.

\subsection{photometry}\label{photometry}
A rigorous photometric calibration is usually performed by applying
a linear transformation on the instrumental (observed) magnitude $m$
of the objects (given by Sextractor Kron-like elliptical aperture fluxes),
\begin{equation} M = m - A_m\times\chi+k_m \times \Delta m+ m_0, \label{eqcal}
\end{equation}
where $M$ is the standard magnitude, $A_m$ is the extinction coefficient, 
$\chi$ is the airmass, $k_m$ is the true-colour coefficient, $\Delta m$ is the
true colour and $m_0$ is the zero-point.

In principle, the coefficient $A_m$ must be derived from the observation of a
field of standard stars at three different airmasses, and the coefficients $k_m$,
from a large range of star colours.  Because the zero-points
$m_0$ and the colour coefficients may vary from CCD chip to CCD chip, one should
also observe the standard field separately in the 8 CCDs of the mosaic at least
three times through the night.

For the UH8K Camera, this task is clearly beyond the observer's reach
because the reading time of the UH8K camera is too long and the whole
night would not be sufficient to observe the fields required for a
proper photometric calibration.  In fact, the dataset provided by the
service observing consists of one field of standard stars SA104
\cite{landolt92}, observed in the middle of the night, in $I$ and
$V$. This minimal observation only allows one to measure an average
zero-point over the whole mosaic in each filter by combining all the
standard stars (one or two per CCD chips).

We derive the colour coefficients $k_V$ and $k_I$ and the magnitude
zero-points $V_0$ and $I_0$ by least-square fit of the colour transformation equations
to the SA104 sequence:
\begin{eqnarray}
(V-m_V)&=&k_V\times(V-I)+V_0\\(I-m_I)&=&k_I\times(V-I)+I_0
\end{eqnarray}
The airmass correction is included in the $m_V, m_I$ magnitudes,
and we use the standard CFHT values for a thick CCD, $A_V=-0.12$ and
$A_I=-0.05$ (http: //www.cfht.hawaii.edu /Instruments /Imaging /FOCAM
/appen.html\#F)
The results appear in Table \ref{table2}. For comparison,
the Table also lists another UH8K measurement obtained 
by J.C. Cuillandre as part of the calibration service of the UH8K 1996 season;
no corresponding errors are provided.

For further control, we also derive the $k_V$ and $k_I$ coefficients
from a synthetic standard sequence: synthetic ``instrumental''
magnitudes are computed for a set of stellar template spectra of
different colours by a routine which uses the theoretical transmission
curve of the instrument optics + filter (Arnouts, priv. comm. 1999).
Application of eq. (3) and (4) to the synthetic sequence yields the
``synthetic'' $k_V$ and $k_I$, also listed in Table \ref{table2}.  The
corresponding zero-points $V_0$ and $I_0$ are derived in a second step
by application of eq. (3) and (4) to all stars of the SA104 sequence,
this time with the ``synthetic'' $k_V$ and $k_I$.

The colour coefficient $k_V$ derived from the synthetic sequence is
compatible with the other values listed in Table \ref{table2}. A
significant dispersion appears in the $k_I$ measurements, probably
because this coefficient is small.  The zero-points $V_0$ and $I_0$
listed in Table \ref{table2} are also consistent within the error bars
and are poorly dependent on the colour coefficients.  Because the
colour coefficients $k_V$ and $k_I$ derived from SA104 display large
errors, we choose to adopt the colour coefficients from the synthetic
sequence and the corresponding zero-points calibrated on SA104:
$k_V=-0.035\pm0.001$ and $k_I=0.005\pm0.0003$; $V_0 = 24.58\pm0.018$
and $I_0 = 24.76\pm0.014$.

For all observed objects in the catalogue, the instrumental magnitudes
are converted into standard $V$ and $I$
magnitudes by re-writing the colour equations eq. (3) and (4), written
in terms of the standard colour $V-I$, into functions of the measured
instrumental colours $(m_V-m_I)$:

\begin{eqnarray} V&=&m_V -A_V \times \chi +
\frac{k_V\times(m_V-m_I)}{1+k_I-k_V}+V_0\label{eq5}\\ I&=&m_I -A_I
\times \chi + \frac{k_I\times(m_V-m_I)}{1+k_I-k_V} + I_0~.
\label{eq6} \end{eqnarray}
$m_I, m_V$ are the observed magnitudes (Kron elliptical apertures),
and $k_V$ $k_I$ $V_0$ and $I_0$ are the values labeled ``synthetic''
in Table \ref{table2}. The other coefficients
have the same meaning as those of eq. (\ref{eqcal}), and we use, as
in eq. (3) and (4), the standard CFH values $A_V=-0.12$ and $A_I=-0.05$.

\begin{figure}
\epsfysize=5cm
\centerline {\epsfbox[50 50 500 500]{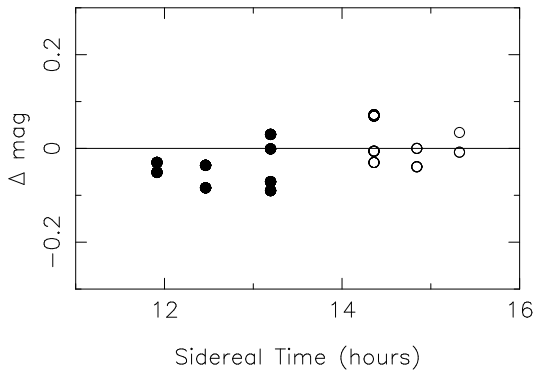}}
\caption{Average magnitude differences $\Delta mag$ of the bright objects
present in the overlapping regions versus the sidereal time. The $I$-band data
are shown as filled symbols, and the $V$-band data as open symbols. Time
variations are always smaller than chip-to-chip zero-point errors.}
\label{figphotvar}
\end{figure}

The astronomer in charge of the service observing stated that the
night was clear with only thin cirrus visible near the horizon at
sunrise (Picat, priv. comm. 1998). The overlapping regions between the
various mosaic fields observed can be used to estimate the possible
variations in the zero-points during the night.  The fields were
observed in the following sequence: UD02-I, UD03-I, UD04-I, UD05-I,
UD05-V, UD04-V, UD03-V, UD02-V. Figure \ref{figphotvar} plots the
average of the magnitude differences $\Delta$mag for the bright
objects (with $I<19$ and $V<20$) detected in the overlapping CCD
regions as a function of sidereal time. For each mosaic field
observed, there are 2 to 4 CCD's presenting an overlap with a CCD
within another field, and each average $\Delta$mag is plotted at the
sidereal time of the first observed overlap.

A small systematic variation of the average $\Delta$mag, denoted
$<\Delta$mag$>$, with sidereal time is detected in the I filter, and
possibly in the V filter.  A field-to-field correction of the zero-points
is done to account for these small variations during the night: we apply to
each field a correction in its zero-point measured by the $<\Delta$mag$>$
in Fig. \ref{figphotvar}, taking the fields UD02-I and UD02-V
as references, and following the sidereal sequence; the
same zero-point correction is applied for all the CCD of each mosaic
field. The residual magnitude variations measured after correction in the CCD
overlaps are $\sigma\simeq0.05$ mag in both the V and I filters. These
put an upper limit on the variations in the zero-points and colour
coefficients between the different CCD's of the mosaic which are not
accounted for in the present analysis.  Note that a gradient remains
in the V data (see section 6.3), which will be removed using the
variation of the average galaxy number counts variations with right ascension.

\begin{figure}
\epsfysize=5cm
\centerline {\epsfbox[0 50 580 580]{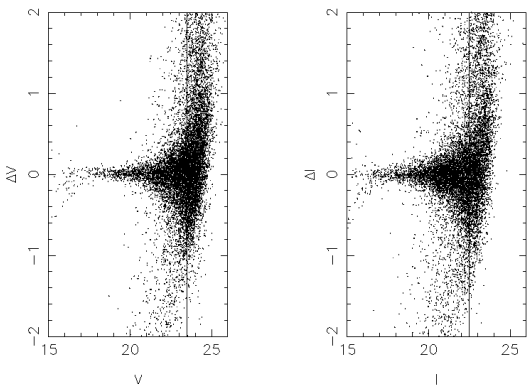}}
\caption{Photometric residuals $\Delta V$ and $\Delta I$ in the overlapping
frames versus magnitude $V$ and $I$. The vertical lines indicate the
completeness limits. The associated standard deviations are
given in Table \ref{table3}.}
\label{fig4}
\end{figure}

To evaluate the final photometric errors in the obtained catalogue, 
the $V$ and $I$ magnitudes of objects in the overlapping sections are 
also compared individually. Figure \ref{fig4} gives the 
residuals in the $V$ and $I$ bands versus magnitude for all objects having
V-I colours (cf section 3.3) and Table \ref{table3} gives the corresponding
standard deviations. The $V-I$ errors are taken to be
$\sim \sqrt{2}~\sigma_{\Delta V}$. The large dispersion at bright magnitudes
in Fig. \ref{fig4} is due to saturation effects. The ``tilted'' variation
of the residuals with magnitude for residuals larger than 0.5 mag in absolute
values (in both the $V$ and $I$ bands) is caused by the following effect: 
each object in the overlaps is given the magnitude measured in one of
the overlap, arbitrarily. If the average of the 2 magnitudes in the 
overlap were used, this ``tilt'' would vanish. However, this affects only 
$\sim 0.5$\% of the objects, and we consider that it would make no significant 
difference in any of the results reported here.
\begin{table}
\caption{Colour coefficients $k$ for the $V$ and $I$ bands derived from the
standard field SA104, J.C. Cuillandre's calibration service, and S. Arnouts'
routine (labeled Synthetic, cf text). The average zero-points
$<$zero-point$>$ were derived using the associated $k$.}\label{table2}
\begin{center}
\begin{tabular}{llll}
\hline
Source&Filter&$~~~~~~k$&$<$zero-point$>$\\
\hline
SA104&$V$&$~~~0.0\pm0.05$&$24.62\pm0.06$\\
SA104&$I$&$~~~0.041\pm0.041$&$24.72\pm0.05$\\
Cuillandre&$V$&$-0.033$&$24.629$\\
Cuillandre&$I$&$-0.066$&$24.721$\\
Synthetic&$V$&$-0.035\pm0.001$&$24.58\pm0.018$\\
Synthetic&$I$&$~~~0.005\pm0.0003$&$24.76\pm0.014$\\
\hline
\end{tabular}
\end{center}
\noindent
\end{table}
\begin{table}
\caption{Photometric errors in $V$ and $I$ (see Fig. \ref{fig4}). The
errors combine random noise, residual variations from field-to-field
in the zero-points during the night, and the uncorrected variations of 
the zero-points and colour coefficients from CCD-to-CCD within the mosaic.}\label{table3}
\scriptsize
\begin{center}
\begin{tabular}{cccc|cccc}
\hline
$V$&$N$&$\sigma_{\Delta V}$&$<\Delta V >$&$I$&$N$&$\sigma_{\Delta I}$&$<\Delta I
 >$\\
\hline
$15.5$&$   2$&$  0.018$&$   0.029$&$15.0$&$  3$&$ 0.030$&$ -0.017$\\
$16.0$&$  14$&$  0.061$&$  -0.021$&$15.5$&$ 23$&$ 0.046$&$  0.037$\\
$16.5$&$  14$&$  0.076$&$  -0.036$&$16.0$&$ 23$&$ 0.061$&$  0.006$\\
$17.0$&$  16$&$  0.067$&$  -0.002$&$16.5$&$ 37$&$ 0.016$&$  0.023$\\
$17.5$&$  27$&$  0.048$&$  -0.013$&$17.0$&$ 45$&$ 0.030$&$  0.014$\\
$18.0$&$  33$&$  0.045$&$   0.019$&$17.5$&$ 51$&$ 0.071$&$  0.025$\\
$18.5$&$  39$&$  0.042$&$   0.010$&$18.0$&$103$&$ 0.027$&$ -0.005$\\
$19.0$&$  69$&$  0.048$&$   0.002$&$18.5$&$108$&$ 0.047$&$  0.011$\\
$19.5$&$ 106$&$  0.060$&$  -0.010$&$19.0$&$203$&$ 0.097$&$ -0.029$\\
$20.0$&$ 138$&$  0.105$&$  -0.000$&$19.5$&$340$&$ 0.052$&$  0.015$\\
$20.5$&$ 201$&$  0.120$&$  -0.011$&$20.0$&$498$&$ 0.122$&$  0.043$\\
$21.0$&$ 321$&$  0.143$&$  -0.017$&$20.5$&$752$&$ 0.101$&$  0.036$\\
$21.5$&$ 502$&$  0.188$&$  -0.030$&$21.0$&$1077$&$ 0.138$&$ 0.063$\\
$22.0$&$ 856$&$  0.233$&$  -0.039$&$21.5$&$1360$&$ 0.157$&$ 0.075$\\
$22.5$&$1244$&$  0.274$&$  -0.032$&$22.0$&$1666$&$ 0.238$&$ 0.072$\\
$23.0$&$2005$&$  0.333$&$  -0.049$&$22.5$&$1921$&$ 0.307$&$ 0.091$\\
$23.5$&$2596$&$  0.379$&$  -0.015$&$23.0$&$1712$&$ 0.407$&$ 0.120$\\
$24.0$&$2247$&$  0.471$&$   0.183$&$23.5$&$1003$&$ 0.505$&$ 0.138$\\
$24.5$&$ 887$&$  0.574$&$   0.398$\\
\hline
\end{tabular}
\end{center}
\noindent
\end{table}
\subsection{Magnitude and colour completeness} \label{completeness}
Three catalogues are generated from the data: one catalogue in the $V$ band,
one in the $I$ band, and one catalogue containing objects with measured $V-I$
obtained by merging the two catalogues. All objects whose centroids are
separated by less than $1\arcsec$ were merged, and their $V-I$ are computed.

The completeness magnitudes of the $V$ and $I$ catalogues are defined
to be one half magnitude brighter than the peak of the distributions. This
corresponds to $V_{complete} \simeq 23.75$ and $I_{complete}\simeq22.75$.
Because some chips go deeper than others and because the correlation
analysis is sensitive to chip-to-chip number density variations, we
lower the completeness limit to the least sensitive chip ($0.2$ magnitude
brighter), to which we add the chip-to-chip dispersion of $\sigma 0.05$  mag
measured from Fig. 4.
Hence, the final completeness limits are $V_{complete} \simeq 23.5$ and
$I_{complete}\simeq22.5$.

\begin{figure}
\epsfysize=5cm
\centerline {\epsfbox[50 50 500 500]{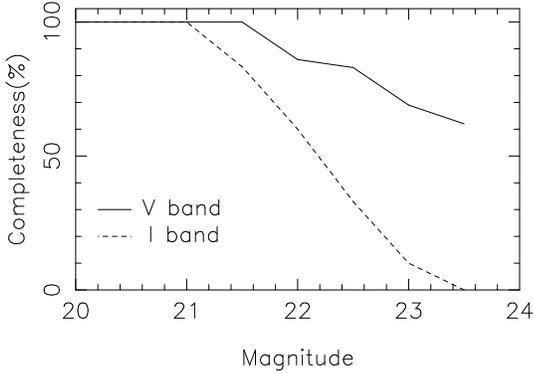}}
\caption{Ratio of galaxies that are detected in the two bands (completeness)
versus $V$ and $I$ magnitudes. At faint $I$ magnitudes, the catalogue is
biased against red objects}
\label{figcolcomp}
\end{figure}

The colour completenesses in the $V$ and $I$ bands are given relative to 
one another in Fig. \ref{figcolcomp}. The red galaxy completeness limit is
determined primarily by how deep the V-band data extend. Because the $V$-band
catalogue is only complete to $V\simeq 23.5$, faint objects redder than $V-I>1$
will be missed near the limit of the $I$ catalogue.  (see Fig. \ref{figgalcol}
and \ref{figgalvimag} in section 4.3). This is an important fact to be
remembered when making colour-selected correlation analyses.
\section{Stars and galaxies}\label{counts}
\subsection{Star/galaxy separation}
\begin{figure}
\epsfysize=7cm
\centerline {\epsfbox[0 40 550 750]{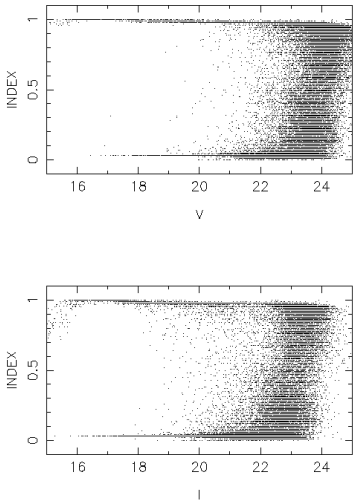}}
\caption{SExtractor star/galaxy classification. Stars have value 1 and galaxies
have value 0. The classification is robust up to $V<22$ and $I<21$}
\label{fig5}
\end{figure}
SExtractor computes a stellarity index for each detected object (in the
interval 0-1, with 1 for stars, and 0 for galaxies). The stellarity index is
determined from a non-linear set of equations (Trained Neural Network)
\cite{bertin96}. The good seeing of the images ($0.7\arcsec-0.8\arcsec$) allows
a robust classification to $V<22$ and $I<21$. According to Bertin \& Arnouts,
the algorithm success rate at these magnitudes is 95\% using data with a
similar sampling and a seeing $\sim 1\arcsec$.
Figure \ref{fig5} shows the index of the $\sim30,000$ objects detected both in 
$V$ and $I$. For $I<21$ or $V<22$, all objects with an
index $<0.9$ are classified as galaxies. This criterion classifies as galaxies
only the objects showing a clear evidence of extendedness. 
For $I>21$ or $V>22$ objects with an index$<0.95$ are classified as
galaxies. Because at these faint magnitudes most objects have an index$<0.95$ 
(the great majority of objects with a stellarity index higher than 0.95 
are spurious detections), the threshold of 0.95 does not remove the remaining 
stars from the sample. To correct for the fact that most of the
stars with $V>22$ have been misclassified as galaxies, we need to apply a
correction for the star dilution (see subsection on data-induced errors in
section \ref{anal}). We evaluated the stellar contamination with the Galaxy
star-count model of Bahcall \& Soneira \cite*{bahcall86}, which is compared
on Figure \ref{fig6} to the number counts of galaxies. The galaxies outnumber 
the stars by nearly an order of magnitude where the classification algorithm 
efficiency is less than 95\% (vertical dotted line in the lower part of the
diagrams). 
\subsection{Galaxy counts}
\begin{figure*}
\epsfysize=10cm
\centerline {\epsfbox[0 0 600 600]{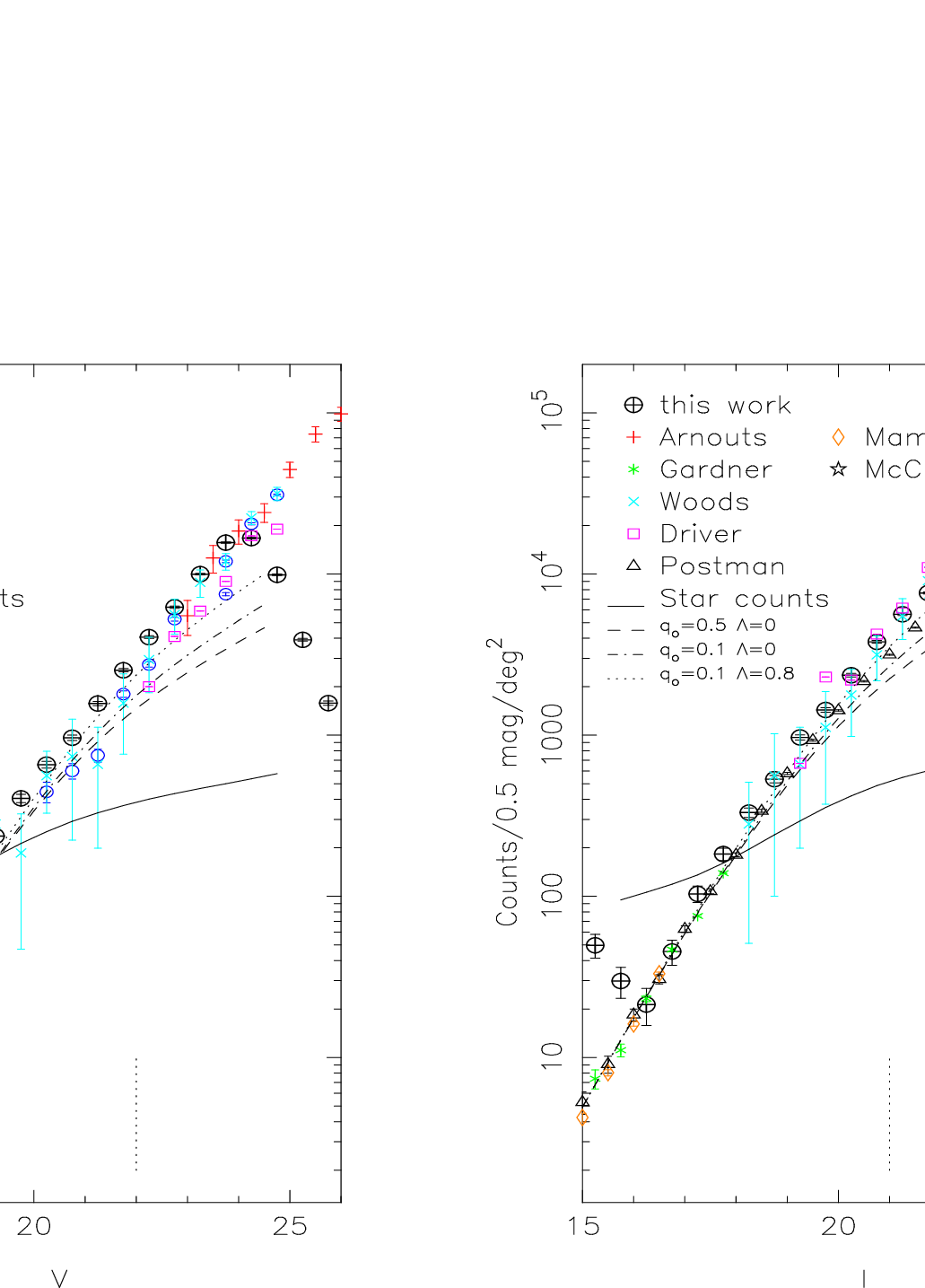}}
\caption{Comparison of our UH8K galaxy counts in $V$ and $I$ with the results of
Arnouts et al. \protect \cite*{arnouts99}, Cowie et al.
\protect \cite*{cowie88}, Drivers \& Phillips \protect \cite*{driver94},
Gardner et al. \protect \cite*{gardner96}, Postman et al. \protect \cite*{post98}, and Woods \& Fahlman
\protect \cite*{woods97}, Mamon \protect\cite*{mamon98}, and 
McCracken et al. \protect \cite*{mccracken00a}. 
Stars counts (solid line) are given according to the Bahcall model
of the Galaxy. The vertical dotted line gives the limiting magnitude below which
the star/galaxy classification is reliable (Fig. \protect \ref{fig5}).
Galaxy number counts are given for three cosmologies in $V$ and $I$ bands (cf text).
The counts are in marginal agreement with a no-evolution flat $\Lambda$ universe.}
\label{fig6}
\end{figure*}
The galaxy counts shown in Fig. \ref{fig6} are in good agreement with other 
measurements. In the $I$ band, We systematically measure $20-30\%$ more galaxies
than Postman et al. \cite*{post98} up to $I<22$. At $I=21.25$, we count 5626
galaxies per deg$^2$, and Postman et al. find 4057. Given the errors in
the $I$ zero-point calibration ($\sigma_{\Delta I}\sim 0.05-0.2$), the possible
difference in the magnitude scale, and the intrinsic cosmic variance, 
we do not consider this difference to be significant.

We model the galaxy counts of Fig. \ref{fig6} following the method described
by Cole, Treyer, \& Silk \cite*{cole92}. They give the equations of
the volume element, the comoving distance and the luminosity distance of the
objects for three cosmologies ($\Omega_0 = 1$, Einstein de Sitter;
$\Omega_0 = 0.2$
Open; $\Omega_0 = 0.2$, $\Lambda = 3(1-\Omega_0)H^2_0$ Flat, a factor $c/H_0$ is missing 
in their equation of the volume element for the flat $\Lambda$ universe). 
Yoshii \cite*{yoshii93b} also details a similar method.
The luminosity function (LF) is chosen to be similar to the CFRS for
which LFs have been measured separately for blue and red galaxies
\cite{lilly95}. 
Here we approximate the LF by its red component. The Schechter
parameterization $\phi(M)$ is used \cite{schechter76},
\begin{eqnarray}
\phi(M)dM&=&0.4\ln10~\phi^*~10^{X(1+\alpha)}dM\times\exp(-10^{X}), \nonumber \\
X&=&0.4(M^* - M), \label{eqlf} \end{eqnarray}
where $M$ is the absolute magnitude in the $V$ or $I$ band, and $\phi^*$,
$M^*$ and $\alpha$ are the Schechter parameters (Here, $M^*_I=-21.5$,
$\phi^*=0.004$, $\alpha=-1.0$). 
$K$ corrections are determined using 13-Gy-old elliptical galaxy template
spectra from the PEGASE atlas \cite{fioc97}, between redshifts $0<z<2$.
As already noted by many authors, this simple model does not
provide a satisfactory fit to the number counts at faint and bright magnitudes
simultaneously for any cosmologies in the $V$-band.
A better fit would include a more realistic luminosity function accounting
for both the red and blue galaxy populations, and for either a density 
or a luminosity evolution (see section 6.1). As our purpose here is not to 
model the number counts, we limit ourselves to this partial model.

Figure \ref{fig6} shows that our UH8K number counts deviate from
the predicted counts in non-evolving Einstein-de Sitter and open
universes at $I\ga21$ and at $V\ga22$.  Postman et al. \cite*{post98}
observed a clear departure from these two cosmological models in their
$I$ counts using the no-evolution model of Ferguson \& Babul
\cite*{ferguson98} (FB).  All other surveys displayed in both panels
of Figure \ref{fig6}
\cite{arnouts99,cowie88,driver94,gardner96,post98,woods97,mamon98,mccracken00a}
have the same behaviour, except for the $V$ number counts of Cowie et
al. \cite*{cowie88}, which agree well with the Einstein-de Sitter
model at $V\le22$ and deviate only at fainter magnitudes.

In contrast, Figure \ref{fig6} shows that the non-evolving flat
$\Lambda$ universe model provides a marginal agreement with our UH8K
galaxy number counts in the $I$ band at our magnitude limit of
$I\simeq22.5$. In fact, most surveys displayed in the right panel of
Fig. \ref{fig6} agree with this cosmological model at $I\la22$, as
expected from the results of the CFRS \cite{lilly96}, where giant red
galaxies evolve little in the redshift interval $0-1$.  In 
Fig. \ref{fig6}, the deviations from the non-evolving flat $\Lambda$
universe occur in dataset which probe the faintest magnitudes, near
$I\simeq 24$. Our sample is not deep enough to show a departure from
this cosmological model. Lidman et al. also find that their $I$ number
counts (not shown in Fig. \ref{fig6}) are compatible with a
no-evolution flat $\Lambda$ universe \cite{lidman96} out to
$I\simeq21$, whereas they deviate from the evolution model of Yoshii
\cite*{yoshii93b} (this model uses a dwarf galaxy blue LF, which
yields very different $K$ corrections at faint magnitudes compared to
a no evolution model). From the mentioned surveys, there are
indications that evolution should be used in the modeling of the $I$
counts at $I>22$, where significant departures from a non-evolving
distribution occur.  

There is also marginal agreement of most $V$ number counts displayed
in the left panel of Fig. \ref{fig6}, including our UH8K data, with a
non-evolving flat $\Lambda$ universe model. The deviations from the
model occur in a wider range of $V$ magnitude, namely at $V\ga22-24$,
depending on the data set, and the deviation is greater than for
the $I$ counts. This may be partly explained by the fact that the $V$
number counts are more sensitive to evolution of the blue galaxy
population than the $I$ counts.

\begin{table}
\begin{center}
\caption{Differential $V$ and $I$-band galaxy counts (see Fig. \ref{fig6}).}
\label{table4}
\begin{tabular}{crr|crr}
\hline
$V$&$N$&$\sigma_N$&$I$&$N$&$\sigma_N$\\
\hline
$  15.75$&$   1.4  $&$   1.4$&$15.75$&$     29.8   $&$  6.5$\\
$  16.25$&$   4.3  $&$   2.5$&$16.25$&$     21.3    $&$  5.5$\\
$  16.75$&$   10.2 $&$   3.8$&$16.75$&$     45.4  $&$  8.0$\\
$  17.25$&$   20.3 $&$   5.4$&$17.25$&$     103.6  $&$  12.1$\\
$  17.75$&$   36.2 $&$   7.2$&$17.75$&$     183.1  $&$  16.1$\\
$  18.25$&$   69.6 $&$   10.0$&$18.25$&$     330.8  $&$  21.7$\\
$  18.75$&$   146.4$&$   14.6$&$18.75$&$     533.9  $&$  27.5$\\
$  19.25$&$   236.3$&$   18.5$&$19.25$&$     968.4  $&$  37.1$\\
$  19.75$&$   406.0$&$   24.3$&$19.75$&$     1434.2 $&$  45.1$\\
$  20.25$&$   656.9$&$   30.9$&$20.25$&$     2355.7 $&$  57.8$\\
$  20.75$&$   964.2$&$   37.4$&$20.75$&$     3794.2 $&$  73.4$\\
$  21.25$&$   1571.8$&$  47.7$&$21.25$&$     5626.0 $&$  89.3$\\
$  21.75$&$   2531.7$&$  60.6$&$21.75$&$     7626.8 $&$  104.0$\\
$  22.25$&$   4058.5$&$  76.7$&$22.25$&$     10312.0$&$  121.0$\\
$  22.75$&$   6219.1$&$  95.0$&$22.75$&$     13972.8$&$  140.8$\\
$  23.25$&$   9991.9$&$  120.4$&$23.25$&$     15936.6$&$  150.4$\\
$  23.75$&$   15665.8$&$ 150.7$&$23.75$&$     12284.4$&$  132.0$\\
$  24.25$&$   16725.8$&$ 155.7$&&&\\
$  24.75$&$   9894.8$&$  119.8$&&&\\
\hline
\end{tabular}
\end{center}
\end{table}
%
\subsection{Galaxy colours}
\begin{figure}
\epsfysize=7cm
\centerline {\epsfbox[0 0 600 600]{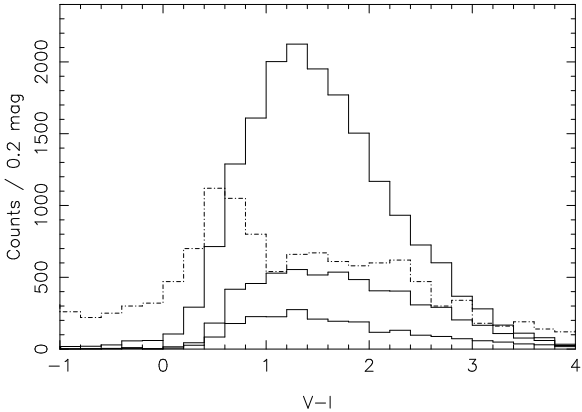}}
\caption{Histogram of galaxy counts (solid line) from top to bottom $I<22.5$,
$I<21$, $I<20$ and star counts (dot-dashed line) vs $V-I$ colour. The star counts
are scaled up by an order of magnitude. Only stars brighter than $I<19$ are
included.}
\label{figgalcol}
\end{figure}

\begin{figure}
\epsfysize=7cm
\centerline {\epsfbox[80 0 680 600]{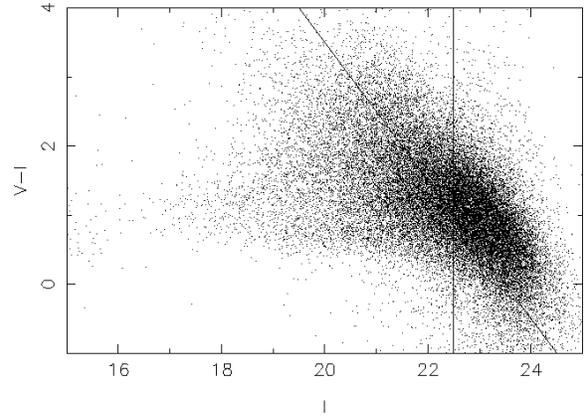}}
\caption{Galaxy $V-I$ colours vs $I$ magnitude. The vertical line indicates the
completeness limit $I=22.5$. The oblique line shows the $V-I$ colours
accessible at the completeness limit $V=23.5$.}
\label{figgalvimag}
\end{figure}

Figure \ref{figgalcol} shows histograms of galaxies and stars versus $V-I$
colours. For galaxies, histograms are given for three limiting magnitudes,
$I<20$, $I<21$,  and $I<22.5$. The corresponding median $V-I$ colours are
$1.45$, $1.57$, and $1.38$. Figure \ref{figgalvimag} shows $V-I$ colours vs
$I$ magnitudes for the sample of $\sim30,000$ detected galaxies. A vertical
line shows the $I$-band completion limit $I=22.5$ and an oblique line shows
$V-I$ colour limits accessible due the $V$-band completion limit of $V=23.5$.
There is no visible trend towards a strong colour evolution of the galaxies
with their magnitudes, but a robust conclusion is not possible because the
$I$ sample is depleted in red objects.
A natural although arbitrary value to divide red galaxies from blue galaxies
is the median of the histogram with $I<22.5$ in Fig. \ref{figgalcol}, located
near $V-I=1.4$. In the rest of the paper red galaxies will be those having
$V-I>1.4$ and blue galaxies will be those having $V-I\le1.4$.\\
\section{Analysis}\label{anal}
\subsection{Estimation of $\omega(\theta)$}\label{LS}
The 2-point angular correlation function $\omega(\theta)$ is calculated
by generating samples of random points covering the same area and having the
same number as the galaxy sample. We use the estimator $W(\theta)$
defined by Landy \& Szalay \cite*{landy93} (hereafter LS), which has the 
advantage of reduced edge effects and smallest possible variance:
\begin{equation}
W(\theta)\equiv W =\frac{DD-2DR+RR}{RR}$~,$ \label{eq7} \end{equation}
where $DD$ is the number of galaxy-galaxy pairs, $DR$ the number of
galaxy-random pairs, and $RR$ is the number of random-random pairs,
all of a given angular separation $\theta$.
Following Roche \cite*{roche96,roche99}, we set a logarithmic binning for the 
separation defined as $\Delta\log(\theta)=0.2$. The numerical approach 
of LS is used to calculate $DD$, $DR$ and $RR$. If one defines the variables 
$d$ and $x$ as
\begin{eqnarray}
d&=&\frac{DD}{G_p(\theta)n(n-1)/2},\\
x&=&\frac{DR}{G_p(\theta)n^2},\label{eq8} 
\end{eqnarray}
\begin{equation}G_p(\theta)=<RR>/[n(n-1)/2],\label{eq10}\end{equation}
then eq. (\ref{eq7}) can be re-written as
\begin{equation}W~=~d~-~2~x~+~1~.\label{eq9}
\end{equation}
$G_p(\theta)$ is the probability of finding two randomly placed galaxies
separated by a distance $\theta\pm d\theta/2$, $n$ is the number of real 
galaxies in the sample, and $<RR>$ is the average over $N$ realizations of
a random sample of $k$ points, with $Nk = 100,000$ ($k$ varies between
$10$ to $100$ depending on the magnitude interval in which the angular 
correlation function is measured).
LS also define $G_t(\theta)$, the probability of finding two neighbors
both at a distance $\theta\pm d\theta/2$ of one given object. 
\begin{equation} G_t(\theta) = <n_t(\theta)> / [n(n-1)(n-2)/2] $,$\label{eq11}
\end{equation} where $<n_t(\theta)>$ is the average number of unique triplets.
$G_t(\theta)$ is necessary to evaluate the random errors (cf section
\ref{error}).
\subsection{Modeling of $\omega(\theta)$, $\xi(r)$ and $r_0$}
The canonical parameterizations of the two-point spatial correlation function 
$\xi(r)$ \cite{phillips78} and of the angular correlation function
$\omega(\theta)$ \cite{peebles80} are\begin{equation}
\xi(r,z)=\frac{(r/r_{00})^{-\gamma}}{(1+z)^{3+\epsilon}}~~~~~$and$~~~~~~
\label{eqxi}~\omega(\theta) = A_{\omega}~\theta^{-\delta} \end{equation}
where $r$ is the comoving distance, $r_{00}$ the correlation length
at $z=0$, $\theta$ the angular separation in radian, $A_\omega$ is the
amplitude of angular correlation function, $\gamma$ and $\delta$ are the slopes
($\delta=\gamma-1$) and $\epsilon$ is a parameter characterizing the
evolution of clustering with the redshift $z$.
If $\epsilon>0$ the clustering evolves in proper coordinates, if $\epsilon=0$ 
the clustering is constant in proper coordinates hence increases in an
expanding universe, if $\epsilon=-1.2$ the clustering is constant in
comoving coordinates. 
The comoving correlation length at a redshift $z$ is related to
$r_{00}$ by
\begin{equation}r_0(z)= r_{00} (1+z)^{-(3+\epsilon-\gamma)/\gamma}.\label{eqr00}
\end{equation}

Given eq. (\ref{eqxi}) and the galaxy redshift 
distribution $N(z)$, one can relate $\omega(\theta)$ and $\xi(r)$: 
\begin{eqnarray} 
\omega(\theta)&=&~C~r_0(z_{peak})^{\gamma}~\theta^{1-\gamma}~B(\epsilon) , \label{eqomega} \\ 
B(\epsilon)&=&\left(\frac{c}{H_0}\right)^{\gamma-1}\int^\infty_0 \frac{r_d(z)^{1-\gamma} N(z)^2}
{g(z)~(1+z)^{3+\epsilon}}~dz\nonumber\\&~&\times\left[\int^\infty_0N(z)dz\right]^{-2},\\
N(z)&=&\left(\pi/180\right)^{-2}\int^{m_2}_{m_1}\sum_i\phi_i(M,z)\frac{dV}{d\omega},\label{eqnz}\\
g(z)&=&\left[(1+z)^2\sqrt{1+\Omega_0 z}\right]^{-1}\quad (\Lambda=0),\\
g(z)&=&\frac{(1+z)^{-1}}{ \sqrt{\Omega_0(1+z)^3-\Omega_0+1}}\quad (\Lambda+\Omega_0 = 1),\\
C&=&\sqrt{\pi}\frac{\Gamma[(\gamma-1)/2]}{\Gamma(\gamma/2)}\simeq 3.68~~~
\label{eqC}. \end{eqnarray}
Here $r_d(z)$ is the angular diameter distance, $dV/d\omega~dz$ is the
comoving volume element, both given for three cosmologies in the appendix
of Cole, Treyer and Silk \cite*{cole92}. $\sum_i\phi_i(M,z)$ is the luminosity
function, whose definition might be dependent on different spectral type $i$
evolving with $z$, and $\Gamma$ is the gamma function.
The value of C is given for a typical value of $\gamma=1.8$ \cite{peebles80}.
Equations (\ref{eqomega}) to (\ref{eqC}) allow to make a direct derivation
of $r_0(z_{peak})$ from $A_{\omega}(\Delta m)$, where $z_{peak}$ if the peak of the
redshift distribution of the galaxies in the sample defined by the interval
of apparent magnitude $\Delta m$:
\begin{equation} 
r_0(z_{peak}) = \left[\frac{A_{\omega}(z_{peak})}{C~B(\epsilon)}\right]^{1/\gamma}.
\label{eqr0}\end{equation}
($A_\omega(\Delta m)$ is rewritten as $A_{\omega}(z_{peak})$). 
\subsection{Estimation of errors} \label{error}
In the past few years, considerable theoretical efforts have been devoted to 
the calculation of errors in the estimation of $\omega(\theta)$. The errors
can be divided into two categories; (1) the random errors; and (2) the systematic 
errors due to the various observational biases in the data. 
The systematic errors are caused by false detections, star/galaxy
misidentifications, photometric variations and astrometric errors. 
The random errors are induced by the finite area of the survey and
depend on the geometry and the size of the sample. 
\subsubsection{Random errors}
Until recently, a Gaussian approximation for the distribution of the
galaxies was assumed to calculate the errors on $\omega(\theta)$. However, the
distribution of the galaxies is known to depart significantly from a Gaussian
distribution on small scales.
A finer approach would be to include the possible correlations 
of the data into the error analysis. This has been done by Bernstein 
\cite*{bernstein94} on the LS estimator $W(\theta)$. Bernstein derives a formal 
solution to the random errors for the case of a hierarchical clustering
universe in the limits $n\gg1$ (number of galaxies in the sample), $W\ll1$,
and $\theta\ll$ angular size of the sample:
\begin{eqnarray} \left(\frac{\Delta W}{W}\right)^2&=&4~(1-2~q_3 + q_4)~
\overline{W}_{\theta_{max}}\nonumber\\&+&\frac{4}{n}\left[
\frac{W_r(1+2~q_3~W)}{W^2}+ q_3-1\right]\nonumber\\&+&\frac{2}{n^2}
\left[(G^{-1}_p -1) \frac{1+W}{W^2} -W^{-1}-1\right]~,\label{eq12}\end{eqnarray}
where the parameters $q_3$ and $q_4$ are measured by Gazta\~naga
\cite*{gaztanaga94}. They are related to the hierarchical amplitudes
$s_3$ and $s_4$ ($s_n \equiv <\omega_n>/[<\omega_2>]^{n-1}$,
where $\omega_n$ are the n-point angular correlation functions),
by $q_3\simeq s_3/3$ and $q_4\simeq s_4/16$ \cite{gaztanaga94}.
$s_3$ and $s_4$ have been derived by Gazta\~naga \cite*{gaztanaga94} and
Roche \& Eales \cite*{roche99} from the APM catalogue \cite{maddox90b}: 
$s_3 \simeq 4$ and $s_4 \simeq 50$  at $\theta \sim 0.1^\circ$, the scale at 
which the angular-correlation function is well measured in the APM catalogue 
\cite{maddox90}; therefore $q_3\simeq 1.3$ and $q_4\simeq 3$.
$W_r$ ($r$ for ring) is the 3-point angular correlation for triplets
of galaxies defined by 2 galaxies at a distance in the interval
[$\theta$,$\theta+\delta \theta$] from the 3rd galaxy. In principle, 
$W_r$ is marginally greater than $W(\theta)$ as defined in eq. (\ref{eq7})
but $W_r\simeq W$ is a fair approximation. $\overline{W}_{\theta_{max}}$
is the average value of the angular correlation function at the largest angular 
separation of the sample. This quantity is not directly accessible because
the estimator $W$ is biased by finite volume errors at separations similar to
the size of the survey. To get an estimation of $\overline{W}_{\theta_{max}}$
we divide our sample into 8 sub-samples (one separation at mid-declination and
three separations in right ascension) and measure the LS estimator $W$.
Then, assuming a power-law whose slope is set at the small angles we extrapolate
the value $W_{\theta_{max}}$ for each sub-sample. Finally, $\overline{W}_{\theta_{max}}$
is obtained as the average over the 8 sub-samples. 
Subdiving the sample into 8 sub-samples
also allows to measure the error on $W$ independently at all scales (the
largest angular separation being a quarter of the largest full survey
separation), by simply taking the variance over the 8 sub-samples. The resulting
variance largely underestimates the variance defined in eq. (\ref{eq12}).

\subsubsection{Systematic errors}
\paragraph{The cosmic bias or integral constraint:}

The measurement of $\overline{W}_{\theta_{max}}$ provides the cosmic bias
$b_{W}$ as 
\begin{equation}b_{W}(\theta)\simeq(3-4~q_3-W(\theta)^{-1})~\overline{W}_{\theta_{max}}
.\label{eq13} \end{equation} 
(Colombi, priv. comm. 1999). This bias can be corrected in the
calculation of $W(\theta)$ in the form of an additive factor $IC$.
At small angular separations eq. (\ref{eq13}) simplifies as
$b_{W}\simeq -\overline{W}_{\theta_{max}}/W$.  This negative
bias is very small but becomes comparable to $W$ when $\theta$ approaches the
angular size of the sample. The usual way of correcting for the cosmic bias
is to fix a slope for the angular correlation function and to find the
constant value of $IC$ which minimizes the $\chi^2$ fitting to the data.
One may point out that the cosmic bias becomes non-negligible only when the
errors $\Delta W/W$ of the LS estimator given by eq. (\ref{eq12}) are large.
The smallest scale at which $b_{W}$ is $\ga 10\%$ is $\theta \ge 3\arcmin$,
which corresponds to the last 2 points in all curves plotted in Figures 12 and
13. In the interval $17<I<21$ for example, $b_W \simeq -0.09$. However, the
random errors at this scale are also significant, $\Delta W/W \simeq 12\%$,
which gives little weight to these points in the least-square fits of $W(\theta)$.
We therefore consider that the cosmic bias has negligible 
impact on our reported slope and correlation amplitude, and we chose
to ignore the cosmic bias in order to avoid introducing a prior 
information on the slope of $W(\theta)$.

\begin{figure}
\epsfysize=7cm
\centerline {\epsfbox[0 0 600 600]{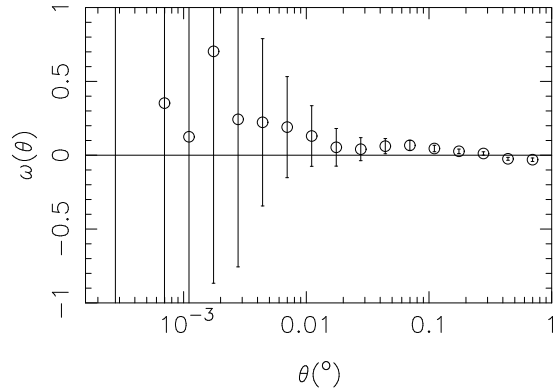}}
\caption{Angular correlation function of stars brighter than $I<21$. The error
bars show 1 $\sigma$ Poisson noise. The result is compatible with a random
distribution.}
\label{fig8}
\end{figure}

\paragraph{Misidentified stars:}
Because stars are uncorrelated on the sky as shown on Fig. \ref{fig8},
the stars fainter than $V>22$ and $I>21$ which are not removed from the
catalogue dilute the clustering present in the galaxy correlation, i.e.
decrease the amplitude of $\omega(\theta)$.
The usual way of correcting for this bias 
\cite{post98,woods97} is to apply a star dilution (multiplicative) factor 
$D_{star}$ to the parameterized amplitude $A_{\omega}$.
\begin{equation} 
D_{star} = \left(\frac{N_{obj}}{N_{obj}-N_{star}}\right)^2~,
\label{eq14} \end{equation}
where $N_{obj}$ is the number of objects in the galaxy samples, 
and $N_{star}$ is the number of stars 
predicted by the Bahcall model of the Galaxy \cite{bahcall86}. For $V<22$ and
$I<21$, $N_{star}$ is given by the classification efficiency of 95\% of
SExtractor, namely the upper value is 5\% of the number of stars detected, so
$D_{star}<1.1$. The values of $D_{star}$ for fainter limiting magnitudes are 
given in Table \ref{table6}.
\paragraph{Masking:}
Although the SExtractor programme is very good at avoiding false detections,
it is sometimes tricked by the diffraction patterns and large wings of bright 
stars as would be any code using an isophotal threshold for detection. Because 
such false detections are strongly clustered, they increase the correlation
amplitude at scales corresponding to the angular size of the false structure.
To prevent the systematic patterns which could be introduced by false
detections, we define two masks, one covering the bad pixels, columns and
regions, and one covering all stars brighter the $V<15$ (from the HST Guide
Star Catalog). 
The second mask is made of four empirical components, (1) a central disk whose
radius is defined by an exponential law as a function of magnitude
($radius=4969\exp[-0.378Mag]$ pixels), (2) a vertical rectangle covering the 
bleeding streak, whose length is an exponential function of magnitude
($bleed=1.28\times10^7\exp[-0.8Mag]$ pixels), (3) and (4) inclined rectangles
(at 38.5$^{\circ}$ and 51$^{\circ}$ of the vertical axis) covering the diffraction
spikes, also following exponential law of the magnitude
($spike=5292\exp[-0.315Mag]$ pixels).
The masked regions are partially visible in Fig. \ref{fig1} because only bright
stars show large wings and the low density of objects does not permit one to
distinguish "true" empty regions from masked regions. We apply the same two
masks to the random realizations for the evaluation of LS estimator $W(\theta)$
in section \ref{LS}.

\paragraph{Photometric errors:} Two kinds of photometric errors may be
present: the random errors or photon noise called $\sigma_{M}$, and
the residual calibration errors in the coefficients $A_V$, $A_I$,
$k_V$, $k_I$ and the zero-points $V_0$ and $I_0$. Note that the errors
given in Table \ref{table3} called $\sigma_{\Delta V}$ and
$\sigma_{\Delta I}$ include both the random and calibration errors.
It is difficult but necessary to evaluate quantitatively the two kinds
of errors because they have opposite qualitative effects on the
angular correlation function.

The random errors can be evaluated from Table \ref{table3}. It is clear
that at bright magnitudes, the calibration errors dominate while random
errors dominate at faint magnitudes.
A random error in a galaxy apparent magnitude is equivalent to an increase of
the possible volume in which that galaxy lies, i.e. it is equivalent to 
a convolution of the de-projected distance interval.  Hence, the random 
error erases the clustering present in the sample.
It is difficult to evaluate directly the decrease in the amplitude
of $\omega(\theta)$ due to the random errors because it is impossible
to disentangle it from a real variation of the spatial galaxy clustering.
Nevertheless, one can follow a simple argument to estimate how random
errors affect the measurement of $\omega(\theta)$. First,
assume that the galaxy number counts follow a power-law 
$log N = \alpha~mag + cnst$ so the relative error is 
$\Delta N/N=(\alpha/\log~e)\Delta mag$.
An extreme case is to consider that all the galaxies with a magnitude error
superior to the magnitude bin for which $\omega(\theta)$ is evaluated 
($\sim 0.5$), are uncorrelated to the sample actually falling in 
the bin. This is an extreme case because many of the galaxies with the large
error in magnitude do belong to the bin. In that case, these galaxies would
have a very similar effect on the amplitude of $\omega(\theta)$ as if they
were stars. So the multiplicative diluting factor $D_{gal}$ of the random
photometric error would be,
\begin{equation} D_{gal}=\left(1-(\alpha/\log~e)\Delta mag\right)^{-2},
\label{eqdgal} \end{equation}
where $\Delta mag$ is the magnitude error and $\alpha$ is the slope of 
the galaxy number count power-law. Taking the best fit slope $\alpha\simeq 0.4$
(see Fig. \ref{fig6} and Table \ref{table4}), a typical magnitude error of 0.15
(see Table \ref{table3}) leads to a dilution factor of 
$D_{gal}\simeq 1.3$, of order of the star dilution factor $D_{star}$ given in Table 
\ref{table6}. This estimate of $D_{gal}$ is an indicative upper limit, and cannot
be used to correct for the dilution due to random errors in the magnitude of
the galaxies because the prior condition is that these galaxies are
uncorrelated. This assumption might not be true, and the correction would
then artificially increase the amplitude of the correlation function.

We now evaluate the calibration error budget. Because
the largest airmass difference in $V$ is $0.267$, and $0.028$ in $I$, even a
$50\%$ error on $A_V$ and $A_I$ would not produce more than $\Delta V\sim0.015$ and
$\Delta I\sim7\times10^{-4}$. If we assume that the chip-to-chip error
on $k_V$ is given by the difference between the values of the synthetic sequence
and the measurement of Cuillandre in Table \ref{table2} (this is certainly not 
the case for $k_I$, because Cuillandre's value is too low) then in the most extreme colours
$(V-I)\sim4$ (only a very small fraction of our sample), the resulting
magnitude difference would be $\Delta V\sim0.008$. The case of $k_I$ is not
clear, although it seems reasonable to assume that the error cannot be more
than ten times the error $\Delta V$, so $\Delta I<0.08$. The residual 
systematic errors in the zero-points have been evaluated in section \ref{photometry}
to be $\sigma \simeq 0.05$. Combining all these mentioned sources of 
calibration errors, one finds a value of 0.053 in $V$ and an upper value of 0.094 in $I$.

The residual photometric errors have an opposite effect on $W(\theta)$.
Namely, they introduces CCD-to-CCD variations in the galaxy number counts 
wrongly interpreted by the correlation analysis as intrinsic clustering on a
scale given by the angular size of the individual CCD chips,
thus artificially increasing the amplitude of $\omega(\theta)$ on these scales.
Geller, de Lapparent \& Kurtz
\cite*{geller84} showed that plate-to-plate systematic variations of more
than 0.05 mag introduced in the Shane-Wirtanen catalogue would produce a
flattening of the correlation function and an artificial break at a scale
corresponding to the plate size. For that reason, we limit our measurement
of the angular correlation to $\theta\la400\arcsec\simeq0.1^\circ$,
corresponding to the smallest dimension of the individual CCD's.
We point out that the
combined effect of random and zero-point errors would be to flatten
artificially the slope of $\omega(\theta)$, as Geller et al. demonstrated.

\paragraph{Astrometric errors:}
Small scale astrometric errors are likely to become significant when the bin
size of $\omega(\theta)$ is of the order of the error. The trivial method to
avoid such contamination is to limit the analysis to bins greater than the
errors, i.e. greater than $0.5\arcsec$ in our case.
When one builds a mosaic survey combining overlapping patches of the
sky, systematic astrometric errors may come into play: the number of objects
may be artificially higher or smaller in overlapping regions just because of
misidentifications. This would induce a similar effect on the amplitude 
of correlation to the zero-point errors at an angular scales of the order
of the field size. However, a comparative analysis 
of the number counts of overlapping regions with other parts of the field
does not show any significant bias toward a higher number of objects in
the overlapping regions.
We therefore consider that our astrometric errors have a negligible effect on 
$\omega(\theta)$, and we limit its calculation to $\theta>1\arcsec$.

\begin{table}
\begin{center}
\caption{Star dilution correction factor $D_{star}$ in $V$ and $I$ bands}
\label{table6}
\begin{tabular}{cc|cc}
\hline
$V$&$D_{star}$&$I$&$D_{star}$\\
\hline
$  22.25$&$   1.08$&$  21.25$&$   1.23$\\
$  22.75$&$   1.11$&$  21.75$&$   1.20$\\
$  23.25$&$   1.10$&$  22.25$&$   1.17$\\
$  23.75$&$   1.09$&$  22.75$&$   1.15$\\
$  24.25$&$   1.08$&$  23.25$&$   1.13$\\
\hline
\end{tabular}
\end{center}
\end{table}
\section{Results}\label{res}
First, we tested the reliability of our code for measuring the LS estimator
of $\omega(\theta)$ on the Zwicky catalogue. The result is consistent
with that of Peebles \cite*{peebles74}. We also used
a deep catalogue (courtesy of Roukema, priv. comm.  1999) and found good
agreement with its $\omega(\theta)$ measurement. 

Figure \ref{fig8} shows
the result of our angular correlation code on the sample of stars brighter
than $I<21$. We limit the calculation of $W(\theta)$ at $0.1^\circ$ because
of the reasons invoked in section 5. The function is always consistent with a
random distribution.
The weak positive signal might be a sign of misclassification of galaxies
or simply small over-densities.
The LS estimator (eq. [\ref{eq9}]) is then measured,
using the entire sample of galaxies, for six cumulative magnitude intervals:
$17<I<20$, $17<I<20.5$, $17<I<21$, $17<I<21.5$, $17<I<22$, and $17<I<22.5$,
and four incremental magnitudes intervals: $20<I<21$, $21<I<21.5$,
$21.5<I<22$, and $22<I<22.5$.

\begin{figure}
\epsfysize=7cm
\centerline {\epsfbox[0 0 550 600]{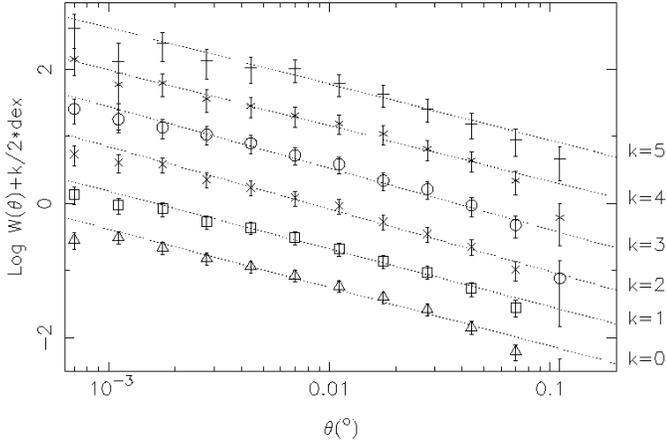}}
\caption{Plots of Log $W(\theta)$ spaced by 0.5 dex (symbols; see also
Table \protect\ref{table7})
and best-fit curves ($A_W$ and $\delta$ of Table \protect\ref{table8})
as dotted line. From top to bottom, $17<I<20$ (k=5),$17<I<20.5$ (k=4),
$17<I<21$ (k=3), $17<I<21.5$ (k=2), $17<I<22$ (k=1), $17<I<22.5$ (k=0)}
\label{fig9}
\end{figure}

Random errors are computed using eq. (\ref{eq12}). Table
\ref{table7} gives the values of $W(\theta)$ and the 3-$\sigma$ random errors
for each cumulative magnitude interval, while Table \ref{table8} gives
the best-fitted values of the corresponding $W(\theta)$ for the parameterization
of eq. (\ref{eqxi}). Three values of the amplitude are given in Table
\ref{table8}. $A_W$ is the best-fit amplitude when both the slope and the
amplitude are fitted. $A^{0.8}_W$ is the best-fitted amplitude when the slope
is constrained to $\delta=0.8$, and  $A^{star}_W$ is $A^{0.8}_W$ corrected
for the star dilution factor ($I>21$) $A^{star}_W=A^{0.8}_W\times D_{star}$.
The amplitudes $A_W$, $A^{0.8}_W$, and $A^{star}_W$ correspond to the
values of the correlation function at $1^\circ$, the usual chosen reference
scale. Table \ref{table8} shows the results for both the integrated intervals
and incremental intervals of magnitudes. All quoted errors are 3-$\sigma$.
For the same $I_{median}$, the values of $A^{0.8}_W$ are consistent between
the integrated intervals and the incremental intervals. Because the error bars
are larger for the incremental intervals, we based our analysis on the 
integrated intervals only. Figure \ref{fig9} plots the LS $W(\theta)$ in 
logarithmic scale for the $I$-band incremental magnitude intervals, 
along with the best-fit power-laws for an
unconstrained slope (Table \ref{table8}). Figure \ref{fig10} plots the
difference of all $W(\theta)$ with the constrained $\delta=0.8$ power-laws.

A final check on the code was done by Colombi (priv. comm. 1999) using a
count-in-cells routine. The results showed that the quality of the dataset
allows to measure higher orders of the distribution of galaxies (skewness \&
kurtosis). The measurements of the angular correlation in the interval
$17<I<21$ gave consistent results with the LS estimator.

Here, we only show the correlation function in the $I$ band. We also measured
$W(\theta)$ on the $V$-band map in the magnitude intervals $17<V<21.5$,
$17<V<22$, $17<V<22.5$, $17<V<23$, and $17<V<23.5$. The values of the
amplitude $A_W$ obtained in the $V$ band are very similar to the values obtained
in $I$, but the average slope $\delta$ is $\simeq0.5-0.6$ in $V$, as compared to
$\simeq0.8-0.9$ in $I$. Neuschaefer \& Windhorst \cite*{neuschaefer95}
also measured slopes $\delta\simeq0.5$ in the $g$ and $r$ bands and
$\delta\simeq0.7-0.8$ in the $i$ band.
\subsection{Variation of slope versus magnitude}\label{slope}
\begin{figure}
\epsfysize=7cm
\centerline {\epsfbox[0 0 550 600]{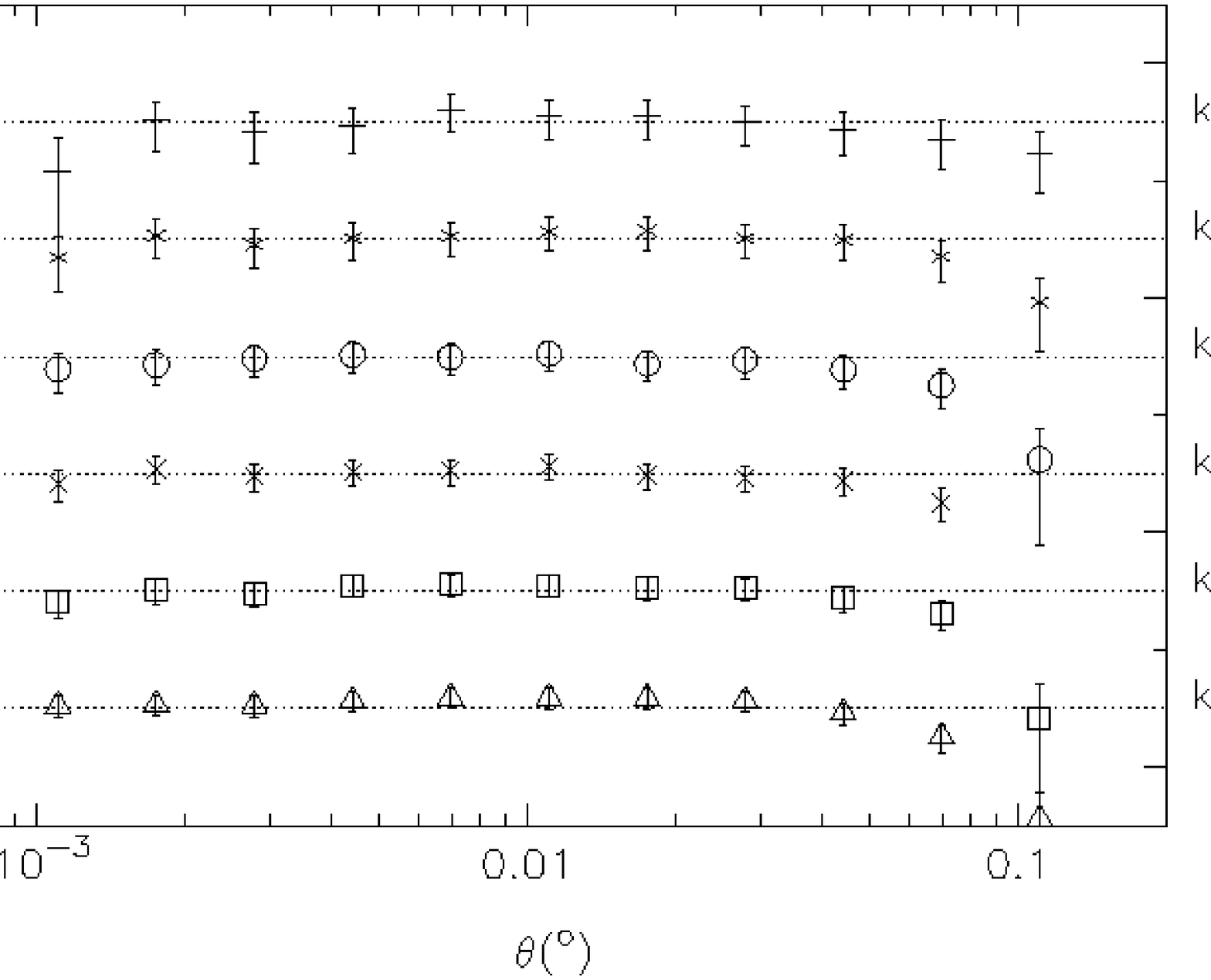}}
\caption{Plots of the differences of the LS $W(\theta)$ and the best-fit
$A^{0.8}_W\theta^{-0.8}$ (dotted lines). The curves are spaced by 1 dex for
sake of clarity. Same magnitude intervals as Fig. \protect\ref{fig9}.}
\label{fig10}
\end{figure}

Recent $\Lambda$CDM models predict a decrease of the spatial correlation
slope $\gamma$ for scales $<10h^{-1}$ Mpc from $\gamma=1.8$ at $z=0$ to
$\gamma\simeq 1.6$ for $z\ge 1$ \cite{kauffmann99b}. This implies a
decrease of the slope $\delta=\gamma-1$ of $W(\theta)$ at small angular
scales and faint magnitudes. Observational evidences are poorly conclusive. 
Brainerd et al. \cite*{brainerd99}
report a steepening of the slope on small scales while
Campos et al{.}, 1995, Neuschaefer and Windhorst, 1995, Infante and Pritchet, 1995,
and Postman et al{.} 1998 find the opposite effect.
Other authors find no significant variations to the limits
of their samples \cite{couch93,roche99,hudon96,woods97}.
Figure \ref{fig10} and Table \ref{table8} show no significant flattening of
slope at faint magnitudes. The slope is compatible with
$\delta=0.8$ for all magnitude intervals. Our result is nevertheless consistent
with the results of Postman et al. \cite*{post98} who find signs of a decrease
only in their faintest bins, at $I>22$, near and beyond our
$I$ limit.

\begin{table*}
\begin{center}
\caption{Measures of $W(\theta)$ and 3-$\sigma$ errors (from eq. [\protect\ref{eq12}]) in the $I$ band for $17<I<20$ ($W_{20}$), $17<I<20.5$ ($W_{20.5}$), $17<I<21$ ($W_{21}$), $17<I<21.5$ ($W_{21.5}$), $17<I<22$ ($W_{22}$), and $17<I<22.5$ ($W_{22.5}$).}
\label{table7}
\begin{tabular}{lrrrrrrrrrrrr}
\hline
$\theta$&$W_{20}$&$3\,\sigma_{20}$&$W_{20.5}$&$3\,\sigma_{20.5}$
&$W_{21}$&$3\,\sigma_{21}$&$W_{21.5}$&$3\,\sigma_{21.5}$
&$W_{22}$&$3\,\sigma_{22}$&$W_{22.5}$&$3\,\sigma_{22.5}$\\
\multicolumn{1}{c}{arcsec}&\multicolumn{12}{c}{$(\times10^{-3})$}\\
\hline
   1.0&-753 &538  &-750 &359 &-1068&242 &-545 &210 &-469 &150 &-377&110\\
   1.6&538  &1041 &-134 &429 &136  &299 &-84.0&176 &-61.1&119 &-222&83.3\\
   2.5&1467 &889  &1346 &574 &760  &309 &554  &194 &422  &124 &291&78.6\\
   4.0&430  &389  &650  &301 &570  &204 &380  &123 &334  &86.8&311&66.9\\
   6.3&767  &356  &561  &216 &446  &143 &392  &105 &287  &66.6&226&46.3\\
  10.0&464  &212  &359  &134 &291  &90.8&236  &63.3&180  &41.6&146&29.8\\
  15.8&372  &152  &272  &95.0&267  &75.5&182  &46.6&136  &30.3&121&23.5\\
  25.1&331  &123  &215  &69.7&171  &48.3&122  &31.1&95.3 &20.9&86.5&16.6\\
  39.8&209  &77.9 &160  &50.3&127  &35.1&94.2 &23.5&71.3 &15.4&63.4&12.2\\
  63.1&145  &54.1 &112  &35.3&71.6 &20.7&56.4 &14.5&45.3 &10.1&41.2&8.06\\
 100.0&83.5 &32.4 &70.1 &22.7&49.7 &14.6&35.3 &9.41&30.2 &6.89&27.5&5.54 \\
 158.5&55.8 &22.2 &43.8 &14.9&26.4 &8.45&21.5 &6.10&17.8 &4.36&17.2&3.65\\
 251.2&30.5 &13.4 &22.0 &8.50&13.3 &4.95&10.3 &3.43&10.7 &2.86&7.09&18.2\\
 398.1&16.4 &8.38 &3.19 &2.99&2.30 &1.93&-1.23&0.51&1.61 &0.95&1.09&6.37\\
 631.0&-2.29&-1.40&-4.12&1.15&-2.27&0.68&-1.80&0.56&-0.48&0.11&-1.44&0.40\\
1000.0&-7.00&-2.18&-5.14&1.62&-2.47&0.96&-0.50&0.16&-0.64&0.31&1.17&0.56\\
1584.9&-3.04&1.88 &-1.85&1.84&-4.98&1.26&-5.00&0.93&-3.68&0.73&-2.50&0.53\\
2511.9&-32  &7.91 &-1.58&2.35&-9.85&0.86&2.27 &1.21&1.15 &0.33&-2.56&0.54\\
\hline
\end{tabular}
\end{center}
\end{table*}
\begin{table*}
\begin{center}
\caption{Best-fit values of the amplitude $A_W$ and the slope $\delta$ of
$W(\theta)=A_W\theta^{-\delta}$; $A^{0.8}_W$ is obtained for a fixed slope $\delta=0.8$
and $A^{star}_W$ is the corresponding star dilution corrected amplitude (see text)
in the interval $10 \arcsec < \theta < 500 \arcsec $. All quoted errors are
$3\,\sigma$. The first 6 rows of the table show the integrated intervals of
magnitudes, the last 4 rows show the incremental intervals of magnitudes.}
\label{table8}
\begin{tabular}{ccccccc}
\hline
\multicolumn{2}{c}{$I$ Magnitude}&$N_{gal}$&$A_W$&$\delta$&$A^{0.8}_W$&$A^{star}_W$\\
Interval&Median&&$\times10^{-4}$&&$\times10^{-4}$&$\times10^{-4}$\\
\hline
17-20.0&19.356&2166&$40.4\pm61.3$&$0.839\pm0.329$&$47.9\pm19.6$&$47.9$\\
17-20.5&19.807&3552&$31.7\pm40.8$&$0.830\pm0.277$&$36.1\pm12.5$&$36.1$\\
17-21.0&20.281&5671&$15.6\pm17.8$&$0.917\pm0.240$&$25.8\pm8.02$&$25.8$\\
17-21.5&20.724&8739&$11.42\pm11.87$&$0.930\pm0.218$&$19.95\pm5.48$&$19.95$\\
17-22.0&21.166&13215&$12.6\pm11.4$&$0.861\pm0.190$&$16.6\pm3.85$&$19.4$\\
17-22.5&21.617&19501&$10.06\pm7.80$&$0.872\pm0.161$&$13.8\pm2.88$&$15.9$\\
\hline
20-21.0&20.610&3504&$9.35\pm14.4$&$0.981\pm0.320$&$20.0\pm8.12$&$20.0$\\
21-21.5&21.270&3068&$6.67\pm14.9$&$0.947\pm0.477$&$12.36\pm7.05$&$15.2$\\
21.5-22&21.772&4476&$10.3\pm17.6$&$0.880\pm0.372$&$14.5\pm5.82$&$17.5$\\
22-22.5&22.267&6283&$11.2\pm14.0$&$0.810\pm0.277$&$11.7\pm4.01$&$13.7$\\
\hline
\end{tabular}
\end{center}
\end{table*}
\subsection{Variation of $A_W$ with magnitude}
The choice of a galaxy luminosity function to model the decrease of $A_W$
with the limiting apparent magnitude is crucial because the behavior
of $A_W$ is sensitive to both the parameterized characteristic absolute magnitude $M^*$
and the slope $\alpha$ (eq. [\ref{eqlf}]).  We choose to use the
luminosity function observed in the CFRS (Lilly et al. 1996), derived from
591 galaxies in the range $0<z<1$, keeping
in mind that such a small sample of galaxies can only provide an indication
of a general trend. Because both the CFRS sample and our sample have been selected
in the $I$ band, we thus limit the possible biases due to the
photometric sample selection.

Lilly et al. divide the CFRS sample into a red population (redder
than an Sbc having rest-frame $[U-V]_{AB}=1.38$; $[U-V]_{AB}\simeq[V-I]_{AB}$ at
$z\sim0.5$, $V_{AB}=V$, and $I_{AB}=I+0.48$) and a blue population
(bluer than an Sbc). The red galaxies show no density or
luminosity evolution in the range $0<z<1$, whereas the blue galaxies
show a luminosity evolution of about 1 magnitude in the same redshift range.
Recent measurements made from the CNOC2 survey \cite{lin99}, on $\sim$2000 
galaxies in the  range $0.1<z<0.7$, confirms the
general observations of the CFRS, although the proposed interpretation is
notably different.
The CNOC2 analysis separates the luminosity evolution from the density
evolution. Early and intermediate (red) galaxies show a small luminosity
evolution in the range $0.1<z<0.7$ ($\Delta M^*\simeq0.5$), whereas late (blue)
galaxies show a clear density evolution with almost no luminosity evolution
in the same redshift range, in apparent contradiction with the CFRS results.

We choose to adapt the CFRS LF to our sample, and we proceed as
follows. Galaxies are separated into two broad spectro-photometric
groups, the E/S0/Sab (called red) and the Sbc/Scd/Irr (called blue),
using the median colour $V-I=1.4$ (see section 4.3).  The red group
has a non-evolving luminosity function with parameters
$\phi^*=0.0148~h^3$ Mpc$^{-3}$, $M^*_I=-21.5+5\log10 h$, and
$\alpha=-0.5$ (eq. [\ref{eqlf}]), and the blue group has a
mild-evolving luminosity function with parameters $\phi^*=0.015 h^3$
Mpc$^{-3}$, $M^*_I=-21.56+(1-e^{-2z}))+5\log10 h$, and $\alpha=-1.07$.
The factor $1-e^{-2z}$ equals 0 for $z=0$ and $\sim 1$ for $z\ge1$ and
mimics the observed smooth brightening of $M^*_I$ with redshift.

\begin{figure}
\epsfysize=7cm
\centerline {\epsfbox[0 0 550 550]{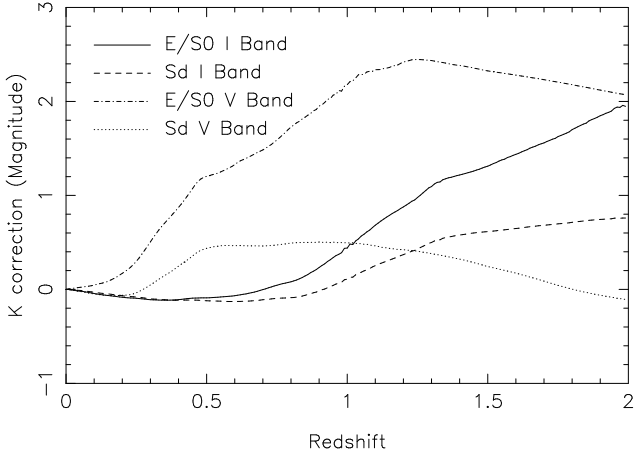}}
\caption{K corrections for early-type and late-type galaxies in $V$ \& $I$ band.
We use the simplest hypothesis of non-evolving galaxy spectra over the
range $0<z<2$.}
\label{figkcorr}
\end{figure}

We integrate eq. (\ref{eqnz}) in the different apparent magnitude
intervals listed in Table \ref{table7}, and obtain $N(z)$ for three
cosmologies.  The $K$ corrections are computed from templates of E (9
Gy) for the red group and Sd (13 Gyr) for the blue group from the
PEGASE atlas \cite{fioc97} and are shown in Fig. \ref{figkcorr}. The
choice of a given atlas is not benign, as Galaz \cite*{galaz98} shows
that $K$ corrections can vary by 50\% at $z\sim 1$ when comparing the
PEGASE atlas with the GISSEL atlas \cite{bruzual93}, leading to
significant differences in $N(z)$.

\begin{figure}
\epsfysize=7cm
\centerline {\epsfbox[0 0 550 550]{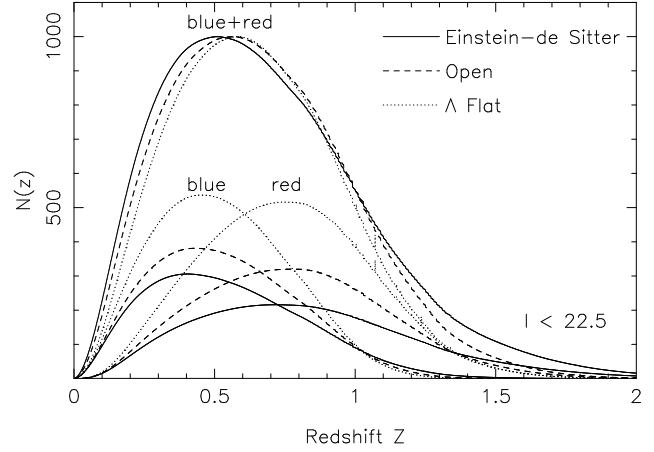}}
\caption{Redshift distributions $N(z)$ for red and blue objects assuming
CFRS-like LF. The limiting magnitude is $I<22.5$. The peaks of the red+blue
distributions are normalized to 1000. The relative number of blue and red 
objects are normalized to the observed ratio in the CFRS survey.}
\label{plotnz}
\end{figure}

Figure \ref{plotnz} shows $N(z)$ for red and blue objects having
$I<22.5$, for three cosmologies: Einstein-de Sitter $\Omega_0=1$ as a
solid line, open universe $\Omega_0=0.2$ as a dashed line, and flat
$\Lambda$ $\Omega_0=0.2$ universe as a dotted line. One can note that
red and blue galaxies show very different redshift distributions as
expected from the different LFs. This should be kept in mind when we
compare the different evolution of $A_W$ with magnitude for the red
and blue samples. The resulting $N(z)$ used to model our sample is the
sum of the $N(z)$ for the red and blue object distributions
respectively. As our 2 colour samples suffer from differential 
incompleteness (see section 3.3), we normalized  the relative number
of blue and red objects to the observed ratio in the CFRS survey at the
corresponding limiting magnitude.

\begin{figure}
\epsfysize=7cm
\centerline {\epsfbox[0 0 550 550]{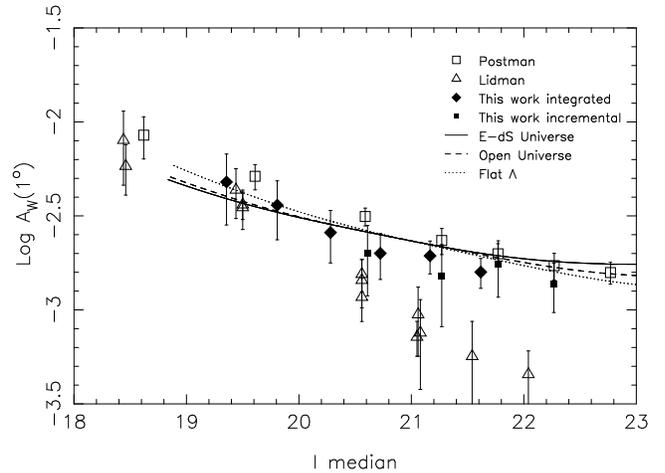}}
\caption{The amplitude of correlation $A^{star}_W$ for the UH8K data is 
plotted against the median $I$ magnitude for the same intervals as in 
Fig. \protect\ref{fig9} and \protect\ref{fig10} (filled symbols) and 
is compared to the results from other groups. The four filled
squares with large error bars are the measurements for the incremental
intervals, and the diamonds show the values for the integrated intervals
(see also Table \protect\ref{table8}). Theoretical curves are calculated
for $\epsilon=-1.2$ and the corresponding best-fit $r_{00}$ (listed in 
Table \protect\ref{tablero}) obtained with $\delta=-0.8$, and for three 
universes: Einstein-de Sitter as solid
lines, $\Omega_0=0.2, \Lambda=0$ as dashed lines and $\Omega_0=0.2, \Lambda$
flat universe as dot-dashed lines.  We use the bimodal CFRS
luminosity function with mild luminosity evolution described in subsection 6.2.}
\label{fig11}
\end{figure}

\begin{figure}
\epsfysize=7cm
\centerline {\epsfbox[0 0 550 550]{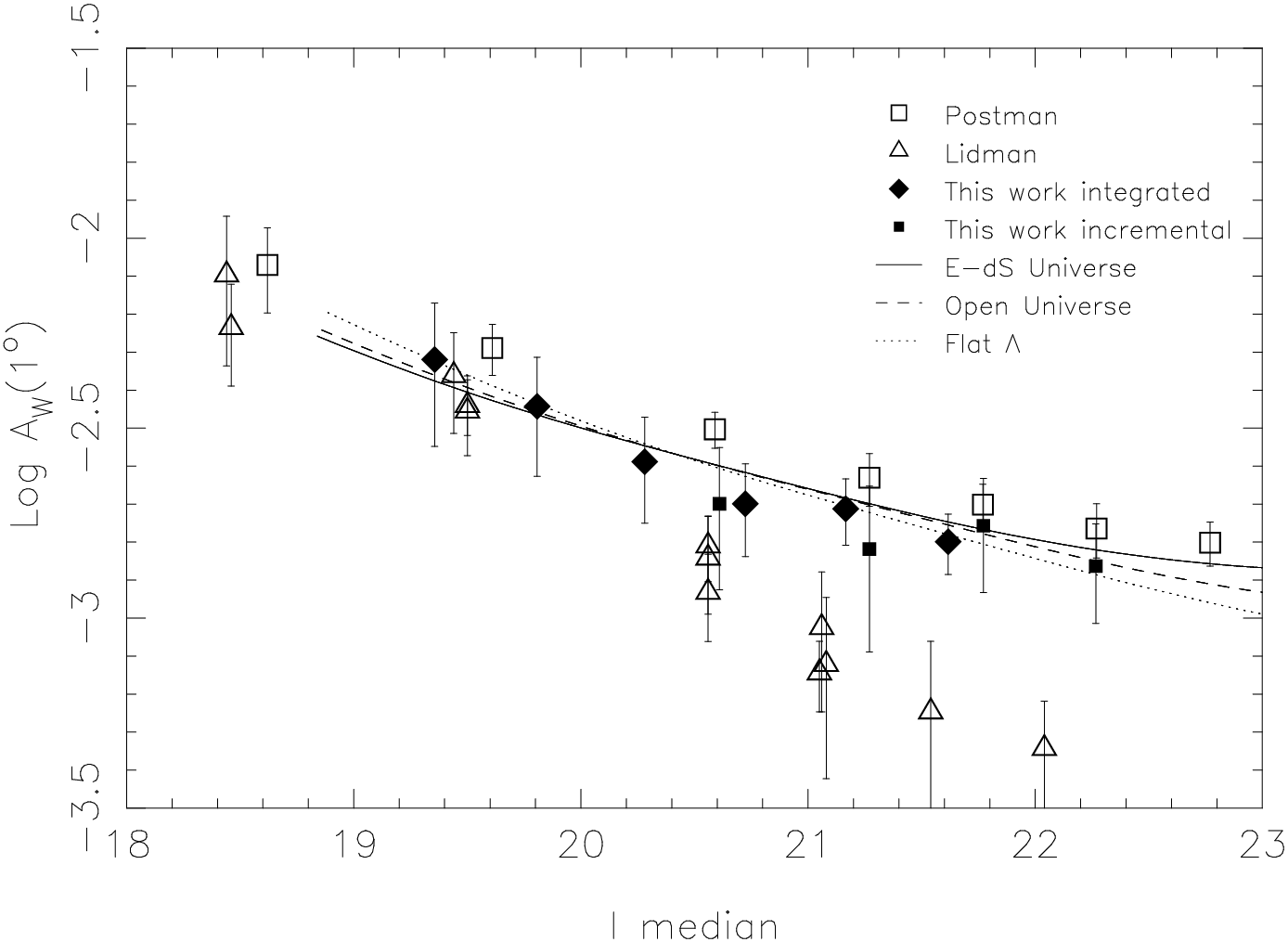}}
\caption{Same as Fig. \protect\ref{fig11}, but with $\epsilon=0$ for the 
theoretical curves. The corresponding best-fit $r_{00}$ are given
in Table \protect\ref{tablero}.}
\label{fig0}
\end{figure}

\begin{figure}
\epsfysize=7cm
\centerline {\epsfbox[0 0 550 550]{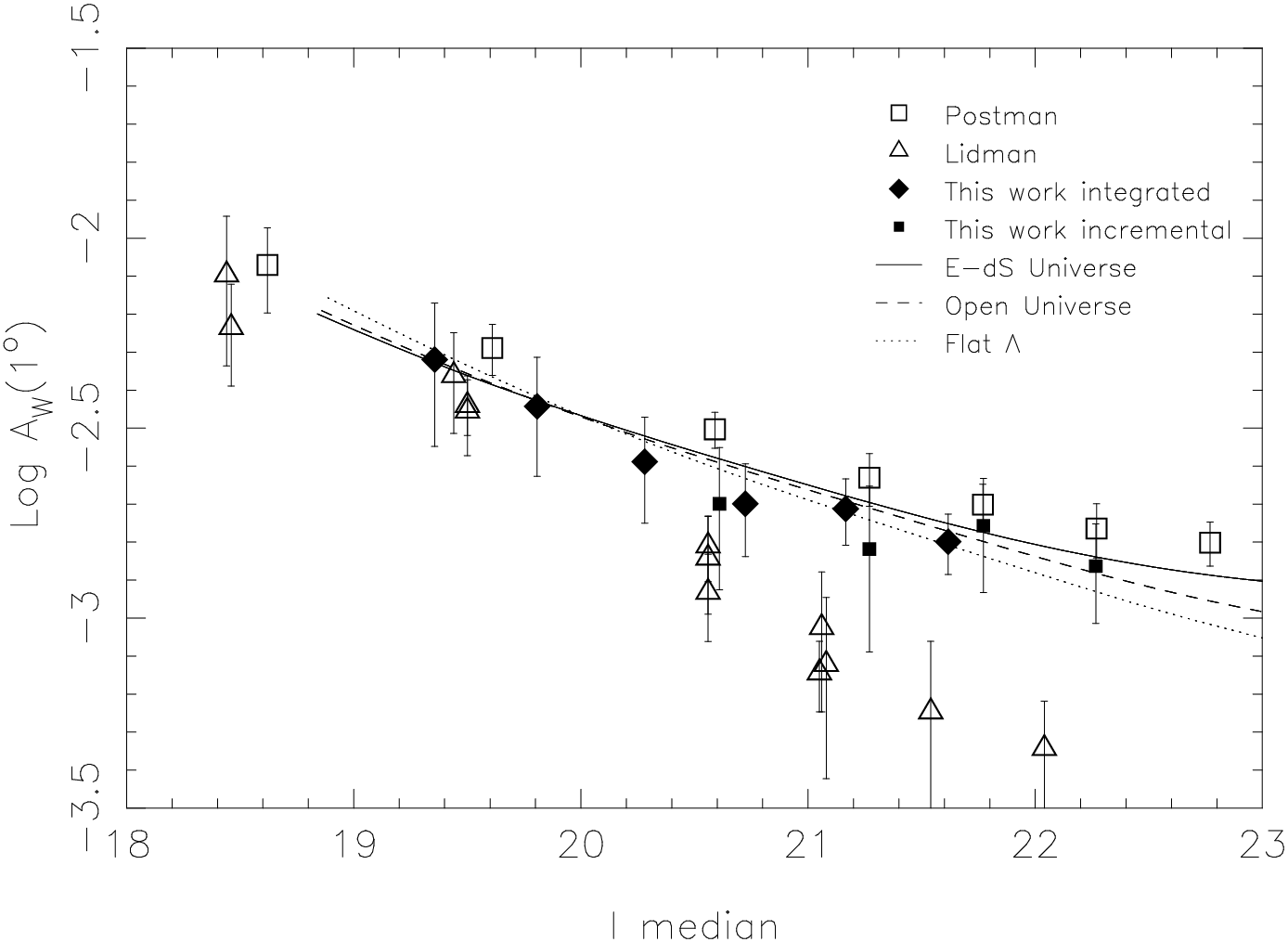}}
\caption{Same as Fig. \protect\ref{fig11}, but with $\epsilon=0.8$ for the 
theoretical curves. The corresponding best-fit $r_{00}$ are given
in Table \protect\ref{tablero}.}
\label{fig0.8}
\end{figure}

Figures \ref{fig11}, \ref{fig0}, and \ref{fig0.8} show the decrease of
the amplitude of the correlation function corrected for star dilution
$A^{star}_W$ for the median $I$ magnitude of the integrated and
incremental intervals (Table \ref{table8}); a fixed slope of
$\delta=0.8$ is used to measure the reference scale, which is taken a
1 degree.  The data-points for the incremental intervals show larger
error bars, but are consistent with the amplitudes for the integrated
intervals.  The measurements of Postman et al. \cite*{post98} and
Lidman \& Peterson \cite*{lidman96} are shown for comparison.  We also
plot the expected curves for the three universes mentioned above, for
each of the three values of the clustering parameter $\epsilon= -1.2$
in Fig. \ref{fig11}, $\epsilon=0$ in Fig. \ref{fig0}, and
$\epsilon=0.8$ in Fig. \ref{fig0.8}). For each value of $\epsilon$,
the theoretical curves are calculated with $\delta=-0.8$ 
and the best-fit spatial correlation length $r_{00}$ at $z=0$, using
eq. (15) and (22).  These values along with the corresponding $\chi^2$
of the fit are listed in Table \ref{tablero} (only the integrated
intervals of magnitude are used for the fitting, 6 data-points). Note that in
Fig. \ref{fig11}, \ref{fig0}, and \ref{fig0.8}, different values of
$r_{00}$ shift the theoretical curves by constant values in the Y direction.

Several conclusions can be drawn from Fig. \ref{fig11}, \ref{fig0},
\ref{fig0.8}, and Table \ref{tablero}.  We can always find a set
$\Omega_0$, $\epsilon$ and $r_0$ which fits our data.  $\Lambda$
universes with $\Omega_m=0.2$ and $\Lambda=0.8$ slightly favour null
$\epsilon$, in good agreement with the results of Baugh et al. (1999)
for semi-analytical models of biased galaxy formation. In contrast,
Table \ref{tablero} shows that Einstein-de Sitter universes favour
positive $\epsilon$ (recall that $\epsilon=-1.2$ means no evolution in
comoving coordinates, and $\epsilon=0$ no evolution in physical
coordinates), as obtained by Hudon \& Lilly \cite*{hudon96}
($\epsilon=0.8$) for the hierarchical clustering CDM model of Davis et
al. \cite*{davis85} in Einstein-de Sitter Universes.  Third, positive
values of evolution parameter $\epsilon$ give better fits to our
observations (Fig. \ref{fig0.8}) than negative $\epsilon$. 
Moreover, comparison of the $r_{00}$ listed in Table \ref{tablero} with
the local values of $r_{00} \simeq 4-8h^{-1}$ Mpc at $z\simeq0$ also
suggest null to mild clustering evolution.

Finally, Table \ref{tablezfid} gives the peak redshift $z_{peak}$
derived from the redshift distributions $N(z)$ (using the CFRS model
luminosity function) for all magnitude intervals and for the three
cosmologies. Using eq. (\ref{eqr0}), the values of $r_0(z_{peak})$ are
computed for all $z_{peak}$, and can be compared to other
results. At $z\simeq0.5$, we measure a value of $r_0(z_{peak})$ in
the range $3.4-3.7\,h^{-1}$ Mpc, depending on the cosmological model
and the evolution index $\epsilon$. The typical 1-$\sigma$ uncertainty
on the values of $r_0(z_{peak})$ listed in Table \ref{tablezfid} is
$\sim0.35\,h^{-1}$ Mpc. The additional error originating from the
uncertainty in the cosmology and in $\epsilon$ is estimated from
Tables \ref{tablero} and \ref{tablezfid} to be $\sim0.35\,h^{-1}$ Mpc.
By adding these errors in quadrature, we find an estimated total error in
the correlation length $r_0$ of $\sim0.5\,h^{-1}$ Mpc.  This
more realistic error can also be applied to the values of $r_{00}$
given in Table \ref{tablero}.

If we assume a flat $\Lambda=0.8$ cosmology, Table \ref{tablezfid}
gives $r_0(z_{peak})\simeq3.5\pm0.5\,h^{-1}$ Mpc at the peak redshift
$z_{peak}\sim0.58$ of the $I\le22.5$ survey. The corresponding value
at $z_{peak}\sim0.50$ is $r_0(z_{peak})\simeq3.7\pm0.5\,h^{-1}$ Mpc.   
Within the error bar, this result is in agreement with most other
angular correlation measurements at $z_{peak}\simeq0.5$
\cite{post98,hudon96,roche99,woods97}.

If we compare with the direct spatial measurements, our result is closer to the
CNOC2 results than those from the CFRS: Carlberg et al. find that the
2300 bright galaxies in the CNOC2 survey \cite{yee96} show little
clustering evolution in the range $0.03<z<0.65$ with
$r_0\simeq3.5-4.5\,h^{-1}$ Mpc at $z\sim0.5$ , depending on the
cosmological parameters \cite{carlberg99}; whereas Le F\`evre et al.
measure $r_0\simeq1.5\,h^{-1}$ Mpc at $z\sim0.5$ for 591 I-selected
galaxies, implying a strong evolution from the local values ($r_0
\simeq 4-8h^{-1}$ Mpc at $z\simeq0$) \cite{lilly95}. As mentioned
above, the small correlation length found in the CFRS survey may be
due to the cosmic variance which affects small area surveys. On the
other hand, neither the CNOC2 survey nor the CFRS survey detect
an evolution in the slope $\gamma\sim1.8$, in good agreement with
our UH8K data.

\begin{table}
\begin{center}
\caption{Best-fit values of the spatial correlation length $r_{00}$ for fixed
$\epsilon$ and three cosmologies.}
\label{tablero}
\begin{tabular}{ccccc}
\hline
$r_{00}$&$\epsilon$&\multicolumn{2}{c}{Cosmology}&$\chi^2$\\
$h^{-1}$ Mpc&&$\Omega_0$&$\Lambda$&\\
\hline
$3.53\pm0.28$&$-1.2$&1&0&0.0115\\
$3.49\pm0.41$&$-1.2$&0.2&0&0.0125\\
$4.11\pm0.21$&$-1.2$&0.2& 0.8&0.00676\\
$4.43\pm0.33$&$0$&1& 0&0.00499\\
$4.38\pm0.28$&$0$&0.2& 0&0.00367\\
$5.03\pm0.28$&$0$&0.2& 0.8&0.00232\\
$5.31\pm0.31$&$0.8$&1& 0&0.00269\\
$5.19\pm0.34$&$0.8$&0.2& 0&0.00218\\
$5.85\pm0.35$&$0.8$&0.2& 0.8&0.00233\\
\hline
\end{tabular}
\end{center}
\end{table}
\begin{table}
\begin{center}
\caption{Redshift $z_{peak}$ of the peak of the redshift
distribution $N(z)$ and the corresponding best-fit correlation lengths
$r_0(z_{peak})$ for different magnitude intervals and cosmologies.
Here $\epsilon=0.8$}
\label{tablezfid}
\scriptsize
\begin{tabular}{ccccc}
\hline
\multicolumn{2}{c}{$I$
Magnitude}&$z_{peak}~~r_0$&$z_{peak}~~r_0$&$z_{peak}~~r_0
$\\
Interval&Median&E.-de S.&Open&Flat $\Lambda$ \\
\hline
17-20.0&19.356&0.295~~~3.98&0.295~~~3.89&0.295~~~4.39\\
17-20.5&19.807&0.325~~~3.88&0.325~~~3.80&0.330~~~4.26\\
17-21.0&20.281&0.355~~~3.78&0.365~~~3.67&0.385~~~4.07\\
17-21.5&20.724&0.400~~~3.65&0.420~~~3.52&0.440~~~3.90\\
17-22.0&21.166&0.455~~~3.50&0.485~~~3.34&0.510~~~3.70\\
17-22.5&21.617&0.505~~~3.37&0.565~~~3.16&0.575~~~3.53\\
\hline
\end{tabular}
\end{center}
\end{table}

\subsection{Variation of $A_W$ with galaxy colour}
\begin{figure}
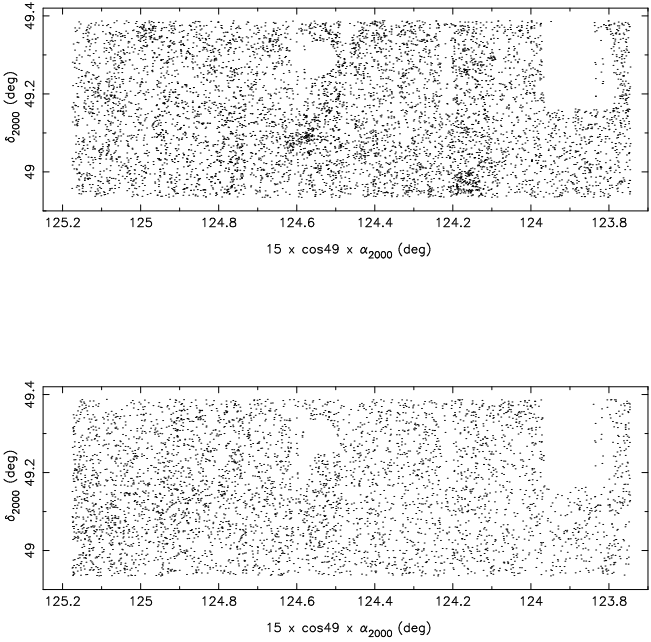

\epsfysize=5cm
\centerline {\epsfbox[0 50 550 450]{9713fig19.ps}}
\epsfysize=5cm
\centerline {\epsfbox[0 50 550 450]{9713fig20.ps}}
\caption{Map of 7069 red galaxies with observed $V-I>1.4$ (top) and
4589 blue galaxies with observed $V-I<1.4$ (bottom) to $I<22$.}
\label{coordblue22.ps}
\end{figure}

We also calculate the variations of $A_W$ with depth for the blue and red
sub-samples of our UH8K data, which are defined in subsection 4.3.
A map of the 8986 red galaxies ($V-I>1.4$) with $I\le22$ is shown in Figure 
\ref{coordblue22.ps} (top panel); the 7259 blue galaxies ($V-I<1.4$) to the same
limiting magnitude are plotted in the bottom panel of Figure \ref{coordblue22.ps}.
One can see that red objects are more clustered than blue objects, as partly
reflected by the morphology density relationship \cite{dressler80}. Note that
the surface density of the blue galaxies increases by a factor
of $\sim2$ with increasing right ascension. This is probably caused by a systematic 
drift in the photometric zero-point along the survey R.A. direction, which was 
not completely removed by the matching of the magnitudes in the CCD overlaps 
(see subsection 3.2).
We then perform an angular correlation analysis on each colour-selected sample.
For the blue sample, we introduce in the random simulations, prior to masking,
the mean R.A. gradient measured from the data.

\begin{figure}
\epsfysize=7cm
\centerline {\epsfbox[0 0 550 550]{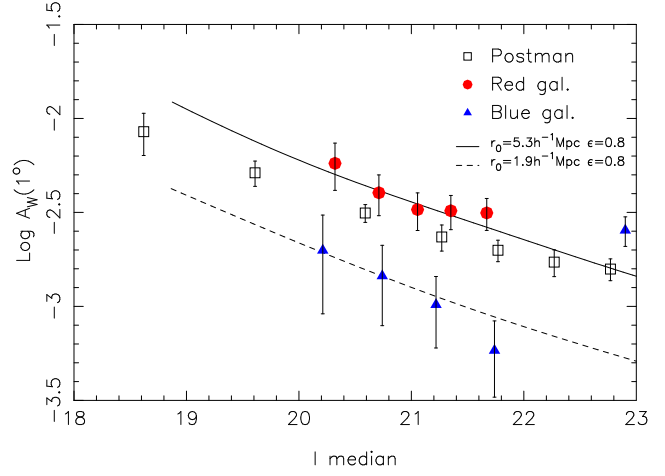}}
\caption{Same as Fig. \protect\ref{fig11}. The galaxies of our sample are
divided into red galaxies with $V-I>1.4$ (filled octagons) and blue galaxies
with $V-I\le1.4$ (filled triangles). The results for the full sample of Postman
et al.
\protect\cite*{post98} (open squares) are shown for reference.
Best-fit models are shown for a $\Lambda$ flat universe with $\epsilon=0.8$
for red galaxies using CFRS red LF (solid line) and
blue galaxies using CFRS blue LF with mild evolution (dashed line).}
\label{figawredblue}
\end{figure}

Figure \ref{figawredblue} shows the decrease of $A_W$ for the red
and blue galaxies to a limiting magnitude $I<22.5$ (see section 3.3 \& 4.3). 
In principle, the separation
should be done on rest-frame colours and not on observed colours, but this
requires prior knowledge of the redshift of the objects. The effect of using
observed colours
is to decrease the resulting angular correlation because galaxies of
different intrinsic colours at different distances are mixed together.
We use the same angular binning as in the previous analyses for consistency
(subsections 6.1 \& 6.2). 
Table \ref{tablemedcol} gives the median magnitude of the red and blue samples.
The last interval given in Table \ref{tablemedcol} includes all 11,483 red
galaxies ($I_{med}=21.670$) and 20,743 blue galaxies ($I_{med}=22.902$) fainter
than $I>17$ (to $V\simeq24$ and $I\simeq23$).

Note that the blue sample reaches fainter $I$ magnitudes than the red sample.
As seen in the section 3.3, this is a selection effect due to the fact that
only galaxies detected in the 2 bands are shown in the sample, and the blue
sample is complete to $V\sim24$. Thus at $I=22.5$, the galaxies redder than
$V-I>1.5$ (in fact most of the red sample) show sparse sampling. 
Each interval in Table \ref{tablemedcol} is respectively complete for
galaxies having $V-I<3, 2.5, 2, 1.5$. We emphasize that the last two intervals
of the red sample ($17<I<22$ and $17<I<22.5$) are probably too incomplete to be 
considered as fair samples.
The incompleteness induces two competing effect. On the one hand, it
increases $A_W$, because the number of objects is artificially small and the
resulting correlation is higher; on the other hand, it makes the
median magnitude brighter (cf Table \ref{tablemedcol}), thus inducing a 
steepening of the slope of the $A_W$ decrease.
In Fig. \ref{figawredblue}, the obviously erroneous offset of
last point for the blue sample, corresponding to the full magnitude
range $I>17$ and a median I magnitude of $\sim22.9$,
also illustrates the error on the measurements of $A_W$
one can expect in extreme cases where magnitudes are poorly defined
and objects are under-sampled.

\begin{table}
\begin{center}
\caption{Median magnitude $I_{median}$ of the red and blue samples for the
different cumulative $I$ magnitude intervals.}
\label{tablemedcol}
\begin{tabular}{lcccc}
\hline
Mag. interval&\# red gal.&$I^{red}_{median}$&\# blue gal.&$I^{blue}_{median}$\\
\hline
17-21&3379&20.212&1854&20.212\\
17-21.5&5057&20.711&2934&20.741\\
17-22&7069&21.055&4589&21.219\\
17-22.5&8986&21.351&7259&21.739\\
\hline
$>$17&11483&21.670&20743&22.902\\
\hline
\end{tabular}
\end{center}
\end{table}

Keeping in mind the warning of the previous paragraphs, and the fact
that the error bars prevent us from drawing strong conclusions, it is
striking to see the different amplitude of $A_W$ for the two
colour samples shown in Fig. \ref{figawredblue}. The larger $A_W$
for red galaxies than for blue galaxies for all $I_{median}$ is in
agreement with the well-known higher clustering amplitude of
early-type galaxies (see for recent spatial measurements in the
Stromlo-APM survey by Loveday et al., 1995), which is partly reflected
by the morphology density relationship \cite{dressler80}.  Using the
first three data-points of the red sample and blue samples in
Fig. \ref{figawredblue}, we calculate that the two distributions are
different at a 4-$\sigma$ level. The corresponding clustering
amplitudes are $r_0(z_{peak})=5.3\pm0.5\,h^{-1}$ Mpc for the red
sample and $r_0(z_{peak})=1.9\pm0.9\,h^{-1}$ Mpc for the blue sample;
both are based on a $\Lambda$ universe with $\epsilon=0.8$. For the
purpose of comparing these values of $r_0(z_{peak})$, the mentioned
errors assume that the correlation functions for the red and blue
galaxies have the same slope $\delta\simeq0.8$. This may not be true
(see Loveday et al. 1995), but our UH8K sample does not allow to
address this issue. We also ignore the systematic effect of the
cosmology which would affect the values of $r_0(z_{peak})$ for the two
populations in the same direction.  The resulting difference in the
measured $r_0(z_{peak})$ for the red and blue populations are at the
3-$\sigma$ level.

Note that if the clustering of the red and blue galaxies were
identical, the larger average distance for the red galaxies compared
to the blue galaxies (see Fig. 15) would yield lower $A_W$ for the red
galaxies than for the blue galaxies. The opposite effect is observed.
Moreover, if the larger number of red galaxies (in the first 4
magnitude intervals in Fig. \ref{figawredblue}) in the red sample
(8986) compared to the blue sample (7259, see Table \ref{tablemedcol})
is caused by large random errors at the limit of the catalogues,
as seen in section 5.3.2, these would tend to dilute 
the $A_W$ for the red sample. The detected increased clustering strength
for the red sample over the blue sample is therefore a lower limit on 
the amplitude difference with colour. 
The apparent flattening of $A_W$ with $I_{median}$ for red galaxies in
Fig. \ref{figawredblue} may be due to the incompleteness of red objects
at faint magnitudes as discussed above.

The difference in clustering amplitudes which we measure for our red and 
blue samples agrees with 
observations by Neuschaefer et al. \cite*{neuschaefer95b},
Lidman \& Peterson \cite*{lidman96} and Roche et al. \cite*{roche96}.
Neuschaefer et al. find that disk-dominated galaxies (blue $V-I$) have
marginally lower $A_W$ than bulge-dominated galaxies (red $V-I$) using HST
multi-colour fields. Similarly, Roche et al. observe a 3$\sigma$ difference in
$A_W$ for a sample divided into objects bluer or redder than $b-r=1.64$.
Lidman \& Peterson see a weak difference between two samples separated by
$V-I=1.5$. Other authors don't see any difference between blue and red-selected
samples, such as Woods and Fahlman \cite*{woods97} for a separation 
of $V-I=1.3$, Brainerd et al. \cite*{brainerd95} for $(g-r)=0.3$, 
Le~F\`evre et al. \cite*{lefevre96} in the spatial correlation length of 
the CFRS for rest-frame $(U-R)_{AB}=1.38$, 
and Infante \& Pritchet \cite*{infante95} for $(J-F)=1$.
In all these cases excepting Infante \& Pritchet, who used photographic
plates, both the number of galaxies and the angular scale of the surveys are
small. It might be possible that in these surveys cosmic variance hides a
weak signal.

\section{Discussion}\label{disc}
\subsection{Limber's formula}

Limber's formula (written here as eq. [15] and [16]) relating $\omega(\theta)$
and $\xi(r)$ is strongly dependent on the shape of the redshift distribution
$N(z)$ which depends on the characteristic absolute magnitude $M^*$ and slope $\alpha$
of the luminosity functions of the different types of galaxies (eq. [\ref{eqnz}]).
Locally, these parameters cover a wide range of values with regard to the
environment and the morphological types of the galaxies
\cite{bingelli88,lin99,marzke98,bromley98,galaz00}. No reasons lead us 
to think that galaxies would show less diversity at higher redshifts.
Recent observations (in UV, IR, X, and $\gamma$) of space-borne observatories
do provide evidence in this direction as well as the discordant measurements
of the LF by the CFRS and CNOC2. Hence, a prior knowledge of the detailed
luminosity functions is necessary to go beyond a phenomenological description
of the clustering of galaxies. 

Our computed $r_0$ is credible only if the correct luminosity
functions have been used in Limber's formula: this would guarantee
that the modeled $N(z)$ is close to the true redshift number
distributions for each galaxy type.  Segregating galaxies into red and
blue samples based on observed colour, as we do here, is also a crude
first step, and should rather be performed using intrinsic colour or,
even better, spectral type (cf section 6.3). These in turn would
require knowledge of the galaxy redshifts. Approximate redshifts
can also be obtained along with spectral type for multi-band photometric 
surveys using photometric redshift techniques
\cite{koo99}.  Because none of the required functions and distributions
are available for our UH8K sample, we emphasize that the reported results
can only be taken as phenomenological, and all the comments on the
deduced clustering must be taken with caution.

\subsection{The cosmological parameters}
Most authors take for granted that different cosmological parameters only
lead to minor differences in the evolution of clustering, compared to the
effects due to the uncertain luminosity function.  Figure \ref{fig11},
to \ref{figawredblue} do show that for a given luminosity function
different cosmological parameters induce different values of $r_0$ and
$\epsilon$. From our models, $r_0$ differs by more than $15\%$ between
the $\Lambda$ flat and Open universes. These differences cannot yet be
distinguished with the present data (up to $z\sim1$). The dispersion between
the different observations (Fig. \ref{fig12}) precludes any derivation of
the cosmological parameters.

Given a CFRS luminosity function, the $\Lambda$ flat universe gives
the better fits to galaxy number counts, and clustering evolution of
our sample ($\epsilon =0-0.8$). This is in agreement with a current
(though controversial) interpretation of recent type Ia supernovae
results \cite{schmidt98,perlmutter99}.  Notwithstanding the numerous
modern methods to measure the cosmological parameters, the present
analysis shows that future surveys containing $10^6$ galaxies with
known luminosity functions per galaxy type and redshift interval to
$z\sim1$ will be required to provide good constraints on cosmological
parameters using this technique. 

\subsection{The evolution of clustering}

\begin{figure}
\epsfysize=7cm
\centerline {\epsfbox[0 0 550 550]{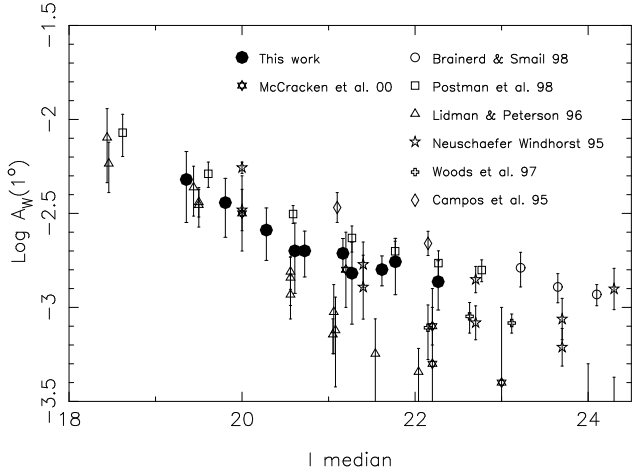}}
\caption{Same as Fig. \protect\ref{fig11}. Our measurements are compared with
recent measurements found in the literature.}
\label{fig12}
\end{figure}

Figure \ref{fig12} shows the decrease of the amplitude of the angular
correlation of our sample compared to most of the other recent
measurements made in the $I$ band
\cite{brainerd98,campos95,woods97,post98,lidman96,neuschaefer95,mccracken00b}.
We applied a magnitude translation between Neuschaefer \& Windhorst $i$
magnitude and our $I$ magnitude of $I=i-0.7$.

In the magnitude range I=20--22, our results are in good agreement
with most of these results except those of Campos et al, which have
the highest amplitude, and on the low side, those of Lidman \&
Peterson and of McCracken et al., 3 times lower in amplitude. 
Postman et al., Woods et al., and Neuschaefer \&
Windhorst obtain intermediate values at $I_{median}>22$.

Postman et al. observe a flattening of $A_W$ for scales $>1'$ and $I>21$. 
The measurements of Brainerd \& Smail extend the flattening of $A_W$ observed by
Postman et al. to $I\sim 24$. We also find a possible flattening in the
decrease of $A_W$ beyond $I \sim 21$.
The most probable explanation for such high dispersion of about a factor ten
between the different measurements of $A_W$ is the dependence of $A_W$ on the
square of the number density. Cosmic variance may account for most
of the discrepancies. The rest may be attributable to systematic errors
of the estimators due to different spatial or magnitude samplings.

As pointed out by Neuschaefer \& Windhorst \cite*{neuschaefer95}, a flattening
of the slope $\delta$ (see eq. [\ref{eqxi}]) with the redshift or with apparent 
magnitude would lower the theoretical curve for $A_W$ 
derived with Limber's formula (eq. [\ref{eqomega}]).
Hence, smaller $A_W$ would still be compatible with a small $\epsilon$,
or a high $\Omega_0$. In other words, clustering would grow faster at smaller
scales than at larger scales. The flattening of the spatial correlation
function is predicted by N-body simulations (Davis 1985; Kauffmann 1999a;
1999b). Neuschaefer \& Windhorst parameterize the flattening of the slope
$\gamma(z)$  with redshift $z$ as: $\gamma(z)=1.75(1+z)^{-C}$,
where $C\simeq0.2\pm0.2$. Postman et al. \cite*{post98} derive $C=0.35\pm0.10$
with a slightly different parameterization: $\gamma(z)=1.8(1+z)^{-C}$.
\section{Conclusions}\label{concl}
In this paper, we present measurements of the angular correlation function for a sample
of $\sim 20,000$ galaxies to $I<22.5$, and $V<23.5$ observed with the CFHT UH8K mosaic
CCD camera over a contiguous area of $\sim30\arcmin\times90\arcmin$.
The main conclusions are the following: 

$\circ$ The amplitude of the angular correlation function of the
complete sample decreases monotonically through the entire range of
magnitude intervals.

$\circ$ The flattening in the decrease of the amplitude, observed by
Postman et al., is marginally confirmed by our analysis.

$\circ$ The best model to fit the evolution of the amplitude of our
sample is the combination of the CFRS luminosity function with mild
luminosity evolution of late-type galaxies and no evolution of
early-type galaxies, a $\Lambda$ flat universe, a clustering evolution
with $\epsilon>0$, and a comoving correlation length of
$r_0\simeq3.7\pm0.5\,h^{-1}$ Mpc at $z\sim0.50$. This in agreement with the
local measurements of $r_0$ with the clustering evolution predicted
by CDM hierarchical clustering models.

$\circ$ Red-selected galaxies show higher amplitudes of correlation than 
blue selected galaxies.

The deep multi-band photometric surveys which are in preparation
should determine whether these observational results on the evolution
of clustering are due to an inadequate definition of the luminosity
functions of the different types of galaxies or whether the actual
clustering differences reflect different formation histories of
disk-dominated vs bulge-dominated galaxies.  Ideally, one would like
to measure the spatial two-point correlation function for each galaxy
type, and for different redshift intervals. The luminosity functions and
their evolution with redshift must be measured accordingly in order to
closely model the observed redshift distribution of the sources.
Application of photometric redshift techniques (Arnouts, 2000 priv. comm.)
to the deep extension of the EIS survey \cite{dacosta99} (http: //www.eso.org 
/science /eis ) should provide new
constraints on these functions. Another deep survey which will also allow to
address these issues is the LZT survey \cite{hickson98}, which will provide
accurate redshifts to $\sigma_z\simeq0.05$ and reliable
spectral types for $\sim 10^6$ galaxies to $z\sim 1$.  This will allow
a more detailed study of the evolution of $A_W$.  Note that the
measured evolution of the clustering amplitude with redshift in these
surveys might also provide useful constraints on the cosmological
parameters.

\begin{acknowledgements}
The authors would like to thank Jean-Pierre Picat who kindly accepted to do the
UH8K observations, Jean-Charles Cuillandre for the UH8K data calibration
service, St\'ephane Colombi for sharing his expertise on the error analysis and
for his count-in-cells routines, B. Roukema for providing his catalogue prior
to publication, S. Arnouts for many discussions, and Gary Mamon for his
careful reading of the manuscript. R.A.C. thanks the Fonds FCAR for a
post-doctoral fellowship.
\end{acknowledgements}
\bibliography{aamnem99,angcorr}
\bibliographystyle{aabib99}
\end{document}